\documentclass[%
 aip,jcp,
 amsmath,amssymb,
 reprint,%
]{revtex4-1}

\usepackage{palatino}
\usepackage[latin1]{inputenc}
\usepackage{epsf}
\usepackage{amsmath,amssymb}
\usepackage{latexsym}
\usepackage{calc}
\usepackage[dvipsnames]{xcolor} 

\usepackage{soul}
\usepackage{bm}
\usepackage[normalem]{ulem}

\usepackage{subcaption}
\usepackage{graphicx}
\PassOptionsToPackage{hyphens}{url}\usepackage{hyperref}
\usepackage{titlecaps}
\usepackage[percent]{overpic} 
\usepackage[export]{adjustbox}

\newcommand{\ben}{\begin{equation}}
\newcommand{\een}{\end{equation}}
\newcommand{\bea}{\begin{eqnarray}}
\newcommand{\eea}{\end{eqnarray}}

\def\ket#1{\vert#1\rangle}

\def\sss{\scriptscriptstyle\rm}

\def\1s{_{1,\sss S}}
\def\2s{_{2,\sss S}}

\def\bR{{\bf R}}

\def\bb{{\bf b}}

\def\bd{{\bf d}}

\draft 

\begin{document}

\title{Exciting DeePMD: Learning excited state energies, forces, and non-adiabatic couplings}

\author{Lucien Dupuy and Neepa T. Maitra}
\email{lucien.dupuy@rutgers.edu; neepa.maitra@rutgers.edu}
\affiliation{Department of Physics, Rutgers University, Newark 07102, New Jersey USA}

\date{\today}

\begin{abstract}
We extend the DeePMD neural network architecture to predict electronic structure properties necessary to perform non-adiabatic dynamics simulations. 
While learning the excited state energies and forces follows a straightforward extension of the DeePMD approach for ground-state energies and forces, how to learn the map between the non-adiabatic coupling vectors (NACV) and the local chemical environment descriptors of DeePMD is less trivial. Most implementations of machine-learning-based non-adiabatic dynamics inherently approximate the NACVs, with an underlying assumption that the  energy-difference-scaled NACVs are conservative fields. We overcome this approximation, implementing the method recently introduced by Richardson [J. Chem. Phys. {\bf 158} 011102 (2023)], which learns the symmetric dyad of the energy-difference-scaled NACV. 
The efficiency and accuracy of our neural network architecture is demonstrated through the example of the  methaniminium cation CH$_2$NH$_2^+$.
\end{abstract}


\maketitle

\section{Introduction}
Since its introduction in 2018, the deep potential methodology~\cite{ZHWCW18,deepmdkitv2} has significantly impacted molecular simulations in physical chemistry, materials physics, and engineering. Even within the past year, we have seen applications in simulation of sodium silicate glasses\cite{bertani2024accurate}, solid-state electrolytes\cite{balyakin2024neural,SSE24}, study of thermodynamic stability of magnesium alloys\cite{HE2023112111}, modeling infrared spectra of liquid H$_2$O\cite{LZHSD24} or ion hydration/exchange at the mineral-water interface\cite{RS24} that would simply not have been computationally feasible otherwise due to the prohibitive cost of ab initio calculations on these high-dimensional and complex systems. 
While several other machine learning methods for energies and force fields exist, an advantage of the the deep potential methodology is its versatility to simulate a wide range of atomistic systems\cite{deepmdkitv2}.

So far DeePMD has largely, but not exclusively~\cite{chen2018deep}, focussed on systems in their electronic ground-state. Extending it to model the wide range of phenomena arising from electronic excitations and ensuing coupled electron-nuclear dynamics, would mean the simulation of processes such as photosynthesis, vision, optoelectronics, photocatalysis, to name just a few, and light-matter interactions in general, could harness the computational advantage of the DeePMD model. With {\it ab initio} methods, the challenge of efficiently modeling excited electronic structure together with the nuclear dynamics tends to limit both the system sizes as well as the time-scales of the simulations, sometimes to the point that the essential points of interest of the phenomena are unattainable. In recent years, different machine-learning methods have enthusiastically stepped in to the non-adiabatic regime to overcome these challenges, at such a pace that already several reviews exist~\cite{LL23, WM21, DB21, LVDL23}. While the earlier developments used kernel ridge regression~\cite{HXLLL18,DBT18}, the community moved more towards neural networks (NNs)~\cite{chen2018deep,westermayr2019machine} to be able to handle large amounts of input data, made available through several softwares~\cite{westermayr2020combining, MLatom24,LRBEBPL21}. { Machine-learning(ML)-powered non-adiabatic dynamics simulations of photo-induced processes enabled the exploration of the nanosecond time scale that would be prohibitively costly with ab initio calculations, such as for CH$_2$NH$_2^+$ photodynamics\cite{westermayr2019machine} or cis-trans isomerization of trans-hexafluoro-2-butene\cite{LRBEBPL21}.} { It also allowed simulations of excited state dynamics simulation of large collections of molecules; for example, an ML model was trained on azobenzene derivatives and then used to predict isomerization quantum yields of thousands of combinatorial species\cite{axelrod2022excited}.}

Most non-adiabatic dynamics simulations use so-called mixed quantum-classical methods, where classical nuclear trajectories are run self-consistently coupling to the electronic system treated quantum-mechanically.
Compared with ground-state processes, these simulations require learning of excited state energies, forces, and non-adiabatic coupling vectors (NACVs). The NACV, also known as the derivative coupling, arises from the nuclear kinetic energy operator acting on the molecular wavefunction, and are defined as:
\begin{equation} \label{eq:NACV}
    \mathbf{d}_{\alpha\beta}(\mathbf{R}) = \langle  \phi^{\alpha}_{\mathbf{R}} |\nabla_{\mathbf{R}}\phi^{\beta}_{\mathbf{R}} \rangle 
\end{equation}
where $\mathbf{R}$ are the nuclear coordinates, and $\phi^{\alpha (\beta)}_{\mathbf{R}}$ is the adiabatic (Born-Oppenheimer) electronic wavefunction associated to the energy $E_{\alpha (\beta)}(\mathbf{R})$. Throughout this paper we will use Greek indices $\alpha,\beta$ to refer only to electronic states.

While learning the excited state energies and forces is more or less a straightforward extension of the ground-state case, learning NACVs is particularly challenging because of three reasons. First, they are typically highly localized, becoming singular at { positions where two or more electronic states are degenerate. Such geometries, named} conical intersections {(CIs), are}  ubiquitous in molecules and prime structures at which electronic population transfer occurs~\cite{schuurman2018dynamics,matsika2011nonadiabatic}. Machine learning methods tend to struggle with very localized quantities compared with smoother ones.
Second, there are sign issues that require additional care due to the arbitrariness of the overall sign of the electronic eigenstate, and due to sign changes induced when encircling a CI\cite{herzberg1963intersection,yarkony1996diabolical,kendrick2002properties}. 
 Third, despite being a vector quantity like the force, the NACV are non-conservative fields, and this makes the imposition of fundamental properties associated with molecular symmetries challenging: For example, while symmetry operations belonging to the point group of the molecule should leave the force unchanged\cite{S91}, which can constrain some of its components to be 0, NACVs do not always obey such constraints. 
 We must retain a rotationally covariant vector prediction for NACV without imposing symmetry constraints they do not obey. To address these challenges, we are in need of different NN designs than those suited for learning a force field.

The techniques that are currently in use~\cite{chen2018deep,  westermayr2019machine,westermayr2020combining} to deal with these three issues  are inherently approximate. The earlier ones  approximate the non-adiabatic coupling from features of the electronic energy surfaces, such as their gaps and gradients~\cite{ZN97}, and are essentially variations of the Landau-Zener approach from the early days of quantum mechanics~\cite{Landau1932,  Zener1932} that are designed to operate within surface-hopping dynamics; these approaches have the additional advantage of bypassing the computation of the NACVs on the training set in the first place. The existing extension of DeePMD to non-adiabatic dynamics~\cite{chen2018deep} utilizes such an approximation. { These methods are not always reliable, as a previous study on several models showed\cite{LZ22}.}  A goal would be to learn electronic structure quantities that could be used for general dynamics methods, not just surface-hopping, especially those with a higher accuracy, e.g.~\cite{BMC99,BM02,WB03,SS12,MAG15,M19b,DRM24}. One approach is to approximate the NACV through the Hessian of the squared energy-gap, as in Ref.~\cite{westermayr2020combining}, which pointed out the computational advantage of machine-learning second-derivatives compared to the {\it ab initio} electronic structure case.
However, it is desirable to go beyond this approximation and try to learn the first-principles expression of Eq.~(\ref{eq:NACV}). 
To this end, Ref.~\cite{westermayr2019machine}  addresses the first challenge in learning Eq.~(\ref{eq:NACV}) by using the Hellmann-Feynman theorem to recast the NAC in terms of a numerator over an energy-difference denominator, since the numerator tends to be smoother function, and phase-correction/invariant algorithms are used  address the second challenge\cite{westermayr2020combining}. { They relied on the SchNet\cite{schutt2017schnet} deep learning architecture as a basis.}
However to address the third challenge, an uncontrolled approximation is used in that work in which the numerator is assumed to be a conservative field. { This is incorrect since it generally has a non-zero curl. Relatedly, non-zero curls of the NACV itself lead to unusual properties, such as the general non-existence of adiabatic-to-diabatic transformations\cite{mead1982conditions} and the Berry phase effect~\cite{yarkony1996diabolical,zhu2016non} (more in Sec.~\ref{sec:symmdyad})}. 

In this work,  we  overcome the challenges above by 
implementing a strategy recently proposed by Richardson~\cite{richardson2023machine} to efficiently and accurately predict the couplings by learning the mapping between their symmetric dyadic and the local chemical environment descriptors of DeePMD. The learned quantity is the {\it bona fide} NACV of Eq.~(\ref{eq:NACV})   with no approximation imposed upon it, so that the machine-learning procedure can make the most of the the data it is trained with.

The manuscript is organized as follows: In Sec.~\ref{sec:NAC} we describe the computation of the NACVs,  describing the conservative field approximation method used in { SchNet}~\cite{westermayr2020combining}, the symmetric dyad decomposition of Ref.~\cite{richardson2023machine}, and our method to build the dyad from DeePMD local descriptors. In Sec.~\ref{sec:DeePMD} we review elements of theory of neural networks in the context of excited state predictions, focussing on the architecture of DeePMD, and the key modifications we made to incorporate the learning of the NACV. In Sec.~\ref{sec:training}
we give details regarding the training of the NN, and Sec.~\ref{sec:results} shows our results on the photodynamics of the methaniminium  cation CH$_2$NH$_2^+$. We present a  conclusion in Sec.~\ref{sec:conc}.

\section{Challenges in learning non-adiabatic coupling vectors}
\label{sec:NAC}
We focus here on attempts to learn the true NACV of Eq.~\ref{eq:NACV} rather than an approximation of it, and we recall the three challenges to compute these that were mentioned in the introduction. 

The peaked nature of the NACV can be readily appreciated from a Hellmann-Feynman recasting of Eq.~(\ref{eq:NACV}), which can be derived by the following argument. 
Noting that $\nabla_\bR\langle \phi^\alpha_\bR\ket{\phi^\beta_\bR} = 0$ from orthonormality, we deduce $\bd_{\alpha\beta} = - \bd_{\beta\alpha}$, and so, from expanding the left-hand-side of $\nabla_\bR \langle\phi^\alpha_\bR\vert H_{\rm BO}\ket{\phi^\beta_\bR} = \delta_{\alpha\beta}\nabla_\bR E_\alpha$ we obtain the equality
\ben
\langle\phi^\alpha_\bR\vert \nabla_\bR H_{\rm BO}\ket{\phi^\beta_\bR} = \delta_{\alpha\beta}\nabla_\bR E_\alpha(\bR) + (E_\beta - E_\alpha)\bd_{\alpha\beta}
\een
leading to 
\begin{equation}
    \mathbf{d}_{\alpha\beta}(\mathbf{R}) = \frac{\mathbf{b}_{\alpha\beta}(\mathbf{R})}{E_\alpha(\mathbf{R}) - E_\beta(\mathbf{R})} \;\;\; \text{for} \;\; \alpha \neq \beta
    \label{eq:NACVHF}
\end{equation}
where 
\ben
\mathbf{b}_{\alpha\beta}(\mathbf{R}) = \langle  \phi^{\alpha}_{\mathbf{R}} \vert\nabla_{\mathbf{R}} H_{\rm{BO}}  \vert\phi^{\beta}_{\mathbf{R}} \rangle\,.
\een

The highly localized nature of the NACV is manifest in the form of Eq.~(\ref{eq:NACVHF}) since energy-levels typically approach each other in localized regions (avoided crossings). More severely, at a CI, the vanishing of the denominator creates a singularity. 
This motivates to separate the learning of the energies and the learning of the numerator, $\mathbf{b}_{\alpha\beta}(\mathbf{R})$ with the idea that the latter is smooth enough for machine-learning to work well. This deals with the first challenge mentioned in the introduction, but the second and third remain. 

Regarding the second challenge, the output from electronic structure codes arbitrarily assign signs to wavefunctions such that the $\bb_{\alpha\beta}(\bR)$ can randomly switch signs for neighbouring $\bR$, creating havoc in the neural net training which leads to spurious oscillations of the machine-learned NACV. 
To account for this, a phase-less loss function for NACVs was introduced in the SchNet method\cite{westermayr2020combining}. For each element of the dataset, this has the form:
\begin{equation} \label{eq:lph}
    L_{\text{NAC}} = \text{min} \Big\{ \sum_{\alpha<\beta}^{n} \Big|\Big|\mathbf{d}^{{\text{Ref}}}_{\alpha\beta} - \mathbf{d}^{\text{ML}}_{\alpha\beta} s_{\alpha} s_{\beta} \Big|\Big|^2 \Big\} , \; s_{\alpha} = \pm 1
\end{equation}
where $n$ is the number of electronic states under consideration.
Since only the combination of wavefunction signs $\{s_{\alpha}\}$ that minimizes the loss function is used, it ensures the ML prediction will converge to a non-oscillating function as NN functions are inherently smooth. 
While this phase-less loss function resolves the difficulty of learning quantities of arbitrary signs, it does not however resolve the problem of the multivalued property of the NACVs due to a CI. We will return to this issue shortly.

\subsection{Conservative field approximation}
\label{sec:cfa}
Regarding the third challenge, the  existing NN codes for non-adiabatic dynamics, SchNet~\cite{westermayr2020combining} and PYRAI$^2$MD~\cite{LRBEBPL21}, that go beyond the Landau-Zener type of approximations for the NACV, learn the numerator of NACVs  in a similar way to the forces associated to each potential energy surface. That is, they set
\begin{equation}\label{eq:FFNACV}
    \mathbf{b}_{\alpha\beta}(\mathbf{R}) \approx \nabla_{\mathbf{R}} \big[ B_{\alpha\beta} \big] \,.
\end{equation}
Here, $B_{\alpha\beta}$ is a fictitious field which is built from the chemical descriptors in the same way as an energy. It thus gives $\mathbf{b}_{\alpha\beta}(\mathbf{R})$ the same symmetries with respect to the nuclear geometry as a force field, under the approximation that $\bb_{\alpha\beta}(\bR)$ is a conservative field.  
To show that in fact $\mathbf{b}_{\alpha\beta}(\mathbf{R})$ is {\it not} a conservative field, we consider its curl. 
When $\alpha\neq\beta$, we have 
\bea
\nonumber
\nabla_\bR \times \bb_{\alpha\beta}(\bR)  &=& \nabla_\bR (E_\beta(\bR) - E_\alpha(\bR)) \times \bd_{\alpha\beta} \\ &+& (E_\beta(\bR) - E_\alpha(\bR)) (\nabla_\bR \times \bd_{\alpha\beta})
\label{eq:curlb}
\eea
There is no reason to expect that either of these terms are zero in general, i.e. even in the absence of a CI, the numerator $\bb_{\alpha\beta}(\bR)$ is non-conservative and cannot be written as the gradient of a scalar function. 
In the rest of this paper we will refer to Eq.~(\ref{eq:FFNACV}) as the ``conservative field approximation" (CFA). 

In passing, we note that, although not an NAC,  when $\alpha = \beta$ the field $ \bb_{\alpha\alpha}(\bR)$ is conservative in the absence of a CI but non-conservative { otherwise }: { by the Hellmann-Feynman theorem, $\bb_{\alpha\alpha} = \nabla_\bR E_\alpha(\bR)$, so}
\ben
\nabla_\bR \times \bb_{\alpha\alpha}(\bR) = \nabla_\bR \times \nabla_\bR E_\alpha(\bR)
\een
{ The right-hand-side would yield zero unless there is a singularity, but a CI provides such a singularity~\cite{herzberg1963intersection,longuet1975intersection,mead1992geometric,yarkony1996diabolical}.
Further, the singularity leads to the field $\bb_{\alpha\alpha}(\bR)$ not returning to itself upon a full circulation around the CI: 
$\gamma_\alpha = \oint \bb_{\alpha\alpha}\cdot d\bR  = \int_S (\nabla_\bR \times \bb_{\alpha\alpha})\cdot dS \neq 0$ (with  $S$ the surface bounded by the circulation),
and thus being multivalued. Note that $\gamma_\alpha$ is related to but not equal to the well-known Berry phase which takes the circulation of the diagonal $\bd_{\alpha\alpha}(\bR) = \langle\phi^\alpha_\bR\vert\nabla\phi^\alpha_\bR\rangle$ instead of that of $\bb_{\alpha\alpha}(\bR)$.
}

{ The presence of a CI also affects the NACVs, causing a multi-valuedness problem. 
 The adiabatic electronic wavefunction, when chosen real, changes sign upon
encircling a CI involving the corresponding adiabatic potential energy surface\cite{herzberg1963intersection}. In this sense, it is multivalued since it does not return to itself in an adiabatic loop around the intersection. It is possible to introduce a R-dependent phase factor to cancel the sign change, which introduces vector potential-like additions to NACVs\cite{mead1992geometric}.
Moreover, the location of where this sign change occurs in configuration space is arbitrary, and if the other state involved in the intersection also is chosen such that its sign change occurs in the same place, then the NACV between them would be single-valued. However, the NACV between any one of these and a third state would still be multivalued since one changes sign and the other doesn't. Ref.~\cite{richardson2023machine} exemplifies this on a model system with similarities to the molecule we consider. The multi-valuedness indicates that the NACV field (and the numerator) is non-conservative, in direct contradiction to the CFA.}

Moving back again to the case of the absence of a CI, then aside from the general unsettling uncontrolled nature of the approximation of Eq.~\ref{eq:FFNACV}, a severe consequence of the CFA is to do with unphysical symmetry constraints: For ML architectures developed to predict force fields, ensuring covariance of the force prediction with respect to rotation and invariance with permutation symmetry is paramount. In principle, molecular symmetries induce physically meaningful constraints to the force, such as setting it to 0 along any direction perpendicular to a symmetry plane. Some ML architectures achieve this property by design. It is the case of DeepMD approach to predict energies, as we will discuss in next Section. It is also the case in SchNet, the energy prediction being a function of atom species and radial atom distances only\cite{schutt2017schnet}. As the force is obtained by differentiating the energy function with respect to infinitesimal atom displacement, use of chain rule shows that a force component will be null when the derivatives of all radial distances with respect to the associated atom displacement are 0. This occurs for planar geometries, as an out-of-plane (infinitesimal) atom displacement does not change distances to first order. However, what is a desirable property for predicting force fields turns out to be a detriment for modeling NACVs, since components orthogonal to a symmetry plane are typically not zero, and should not be set to 0 due to the design of the NN. We will illustrate this in Sec.~\ref{sec:results}.  In situation when out-of-plane motion is funneling electron population transfer, constraining out-of-plane NACV components to 0 would lead to a significant error on the population dynamics.

\subsection{Symmetric dyad matrix}
\label{sec:symmdyad}
We will implement the approach of Richardson which, while motivated in the original work~\cite{richardson2023machine}  by confronting the multivalued character of the NACV, in fact also overcomes all the challenges above. The central object is an auxiliary quantity, the symmetric dyad of $\mathbf{b}_{\alpha\beta}(\mathbf{R})$, and it allows us to learn NACVs more rigorously keeping with its original definition~\cite{richardson2023machine}.

The symmetric dyad of $\mathbf{b}_{\alpha\beta}(\mathbf{R})$ is defined as \cite{richardson2023machine}:
\begin{equation}
    \Gamma_{\alpha\beta}(\mathbf{R}) = \mathbf{b}_{\alpha\beta}(\mathbf{R}) \; \big(\mathbf{b}_{\alpha\beta}(\mathbf{R})\big)^T
    \label{eq:Gamma}
\end{equation}
As a real symmetric matrix $\Gamma_{\alpha\beta} \in \mathbb{R}^{3 N \times 3 N}$, it has  only one non-zero eigenvalue $\lambda=||\mathbf{b}_{\alpha\beta}(\mathbf{R})||^2$ associated to the eigenvector $\mathbf{b}_{\alpha\beta}(\mathbf{R})/\sqrt{\lambda}$. This makes recovering the NACVs from their symmetric dyad straightforward.

By learning $\Gamma_{\alpha\beta}$, there is no inherent approximation as there is with the CFA. 
Two further advantages of this auxiliary field are its single-valuedness and its independence to phase factors in electronic wavefunctions. During a dynamical simulation, phase tracking along trajectory allows to retrieve the phase accumulated when encircling a CI, thus recovering Berry phase effects lost in the CFA. We return to the specific details of how we learn $\Gamma_{\alpha\beta}$ in Sec.~\ref{sec:learningdyad}.

\section{DeePMD Neural Network architecture}
\label{sec:DeePMD}
Over the years the DeePMD community has developed a great variety of schemes to encode chemical information~\cite{wang2022tungsten,deepmdkitv2}. Here, we focus our discussion only on approaches that either are used in the present work or are relevant to predict vector properties. We encourage the reader to consult Ref.~\cite{deepmdkitv2} for an overview.
We start this section with the building blocks of DeePMD's efficient prediction of electronic structure properties from the local chemical environment.

We distinguish the vector $\mathbf{R}$ containing all nuclear coordinates from $\mathbf{r}_i=(x_i,y_i,z_i)$ containing only the cartesian coordinates of atom $i$.

\subsection{Embedding}

DeePMD relies on the fundamental idea that electronic properties should be obtainable as a sum of atom-wise contributions that only depend on atomic interactions between neighbors, up to a certain distance. To each atom $i$ is associated its own descriptor encoding only the position of neighboring atoms up to a cut-off radius $r_c$ (see below).
All descriptors considered in this work are built upon a smoothly-decaying coordinate matrix of neighboring atoms with rows~\cite{ZHWSCE18}:
\begin{equation} \label{eq:GenR}
    \left( \mathcal{R}^i \right)_j = \left\{ s(r_{ij}) \quad \frac{s(r_{ij})x_{ij}}{r_{ij}} \quad \frac{s(r_{ij})y_{ij}}{r_{ij}} \quad \frac{s(r_{ij})z_{ij}}{r_{ij}} \right\}
\end{equation}
where $\mathbf{r}_{ij} = \mathbf{r}_j - \mathbf{r}_i = (x_{ij},y_{ij},z_{ij})$ and $r_{ij} = ||\mathbf{r}_{ij}||$. 
Thus $\mathcal{R}^i \in \mathbb{R}^{N_c \times 4}$ where $N_c$ is the expected maximum number of neighbors within the cut-off radius $r_c$.
The role of the switching function $s(r)$ is to control the ability of neighboring atoms to influence each other by smoothly transitioning from an inverse distance function to a fast-decaying polynomial depending on cut-off hyper-parameters tweaked by the user: 
\begin{equation}
    s(r) = \left\{ \begin{tabular}{l l}
    $\displaystyle \frac{1}{r}$ & \; $r<r_s$,
    \\ \\
    $\displaystyle\frac{1}{r} \left[ q^3 (-6q^2+15q-10) +1 \right]$ & \; $r_s \leq r < r_c$,
    \\ \\
    $0$ & \; $r \geq r_c $
    \end{tabular} \right.
\end{equation}
where $\displaystyle q=\frac{r-r_s}{r_c-r_s}$  and $r_s$ is a parameter controlling the smooth turn-off within the local region. The design of the switching function is explained in Refs.~\cite{ZHWSCE18,deepmdkitv2}. As it is continuous up to the second-order derivative, $s(r)$ is sufficiently smooth to be used in descriptors that in turn will feed fitting networks predicting energies and forces.

The generalized coordinate matrices $\mathcal{R}^i$ are invariant with respect to system translation, but lack the invariance with respect to rotation. Thus, they are not \textit{bona fide} descriptors, but will constitute their building blocks. 
One of the strengths of DeePMD is to include a machine learning component in the design of descriptors in complement to the fitting networks they will be fed into. Different descriptor types are distinguished by how much information is fed to this embedding neural network.

 The so-called 2-body embedding descriptor only feeds radial information to the embedding NN, denoted $\mathcal{N}_{e,2}$:
\begin{equation} 
\label{eq:2bG}
    \left( \mathcal{G}^i \right)_j = \mathcal{N}^{p_{ij}}_{e,2}\big(s(r_{ij})\big)
\end{equation}
where it is understood that each line of $\mathcal{G}^i \in \mathbb{R}^{N_c \times M}$  is obtained by feeding one scalar value $s(r_{ij})$ to the embedding NN $\mathcal{N}_{e,2}$, returning values of its final neuron layer of width $M$. A specific set of parameters (weights and biases) for $\mathcal{N}_{e,2}$ is used depending on the atomic species of the atom pair ${p_{ij}}$. This  ensures that the descriptor satisfies permutation symmetry, and reduces complexity. The full embedding descriptor of atom $i$ is built following:
\begin{equation}
\label{eq:2bD}
    \mathcal{D}^i = \frac{1}{N_c^2} \left( \mathcal{G}^i \right)^T \theta^i \; \mathcal{G}^i_<
\end{equation}
where $\theta^i = \left( \mathcal{R}^i \right) \left( \mathcal{R}^i \right)^T$ encodes the angular information ('angle form' of Ref.~\cite{deepmdkitv2} ) of two neighbors $j$ and $k$ 
and with $\mathcal{G}^i_< \in \mathbb{R}^{N_c \times M_<}$ taking only the first $M_<$ columns of $\mathcal{G}^i$ to reduce the size of $\mathcal{D}^i \in \mathbb{R}^{M \times M_<}$. We emphasize that formally, the features of this descriptor encode information beyond 2-body terms. The name only refers to the amount of information that is fed to the embedding NN. By design, this descriptor is invariant under permutation symmetry and overall translations/rotations of the system. { By studying its building parts, we can show that when used within the CFA, this descriptor will by design set out-of-plane vector components to 0 for planar geometries. First, as the embedding NN $\mathcal{N}_{e,2}$ is only fed with radial information, it has the same properties as SchNet NN when atoms lie in a plane: It will have strictly 0 derivatives with respect to out-of plane displacement. Second, elements of $\theta^{i}$ are scalar product of two rows of $\mathcal{R}^i$ given in Eq. \eqref{eq:GenR}. These scalar products of position vectors of atoms also have 0 derivatives with respect to out-of-plane atomic displacement for planar geometries. By chain rule, as all parts of the descriptor have null derivatives in this situation, so does $\mathcal{D}^i$ as a whole.}

Better accuracy can be achieved when 3-body terms are fed to the embedding NN \cite{wang2022tungsten}, denoted $\mathcal{N}_{e,3}$.
In this case, tensor $\mathcal{G}^i \in \mathbb{R}^{N_c\times N_c \times M}$ contains embedding NN features obtained as \footnote{We use the definition of Ref.~\cite{deepmdkitv2}}:
\begin{equation}
\label{eq:3bG}
    \left( \mathcal{G}^i \right)_{jk} = \mathcal{N}_{e,3}^{p_{jk}}\left( (\theta^i)_{jk} \right)
\end{equation}
where it is understood that the neural network is separately fed with each scalar $(\theta^i)_{jk}$ and outputs a vector of $M$ features where $M$ is the width of the final neuron layer. A specific set of NN parameters is used for every pair of atomic species $p_{jk}$, similar to what is done for 2-body embedding. 
The full embedding descriptor of atom $i$ reads:
\begin{equation}
\label{eq:3bD}
    \mathcal{D}^i = \frac{1}{N_c^2} \, \theta^i : \mathcal{G}^i
\end{equation}
The notation "$:$" represents the contraction between matrix $\mathcal{R}^i \left( \mathcal{R}^i \right)^T$ and the first two dimensions of tensor $\mathcal{G}^i$. As desired, this descriptor is invariant under permutation symmetry and overall rotation/translation of all atoms. { Having already analysed how its building parts behave w.r.t. out-of-plane displacement for planar geometries, we can conclude it will also strictly set out-of plane vector components to 0 for planar geometries when used together with the CFA.}

We now recall how energies and forces are predicted from those descriptors.

\subsection{Property prediction: energies and forces}
\label{sec:property}

A separate energy predictor will be associated to each electronic state $\alpha$. We describe its  structure for one state only without loss of generality.
Following the principle of atomic embedding, each atomic descriptor $\mathcal{D}^i$ is fed to a fitting network $\mathcal{F}_\alpha$ outputing a single scalar (zero-order tensor) interpreted as an atom-wise energy contribution. The total energy is recovered as the sum over atoms:
\begin{equation}
    E_\alpha = \sum_{i=1}^{N} E_{\alpha,i} = \sum_{i=1}^{N} \mathcal{F}_\alpha^{a_i} (\mathcal{D}^i)
\end{equation}
where the $a_i$ superscript in $\mathcal{F}^{a_i}_0$ indicates that atoms of the same species $a_i$ share the same parameters in the fitting network.
In turn, the force is recovered through automatic differentiation of the NN prediction of $E_\alpha$ with respect to atomic positions:
\begin{equation}
    \mathbf{F}_\alpha = - \nabla_{\mathbf{R}} E_\alpha
\end{equation}

\begin{figure}[h!]
    \centering
    \includegraphics[width=.5\textwidth]{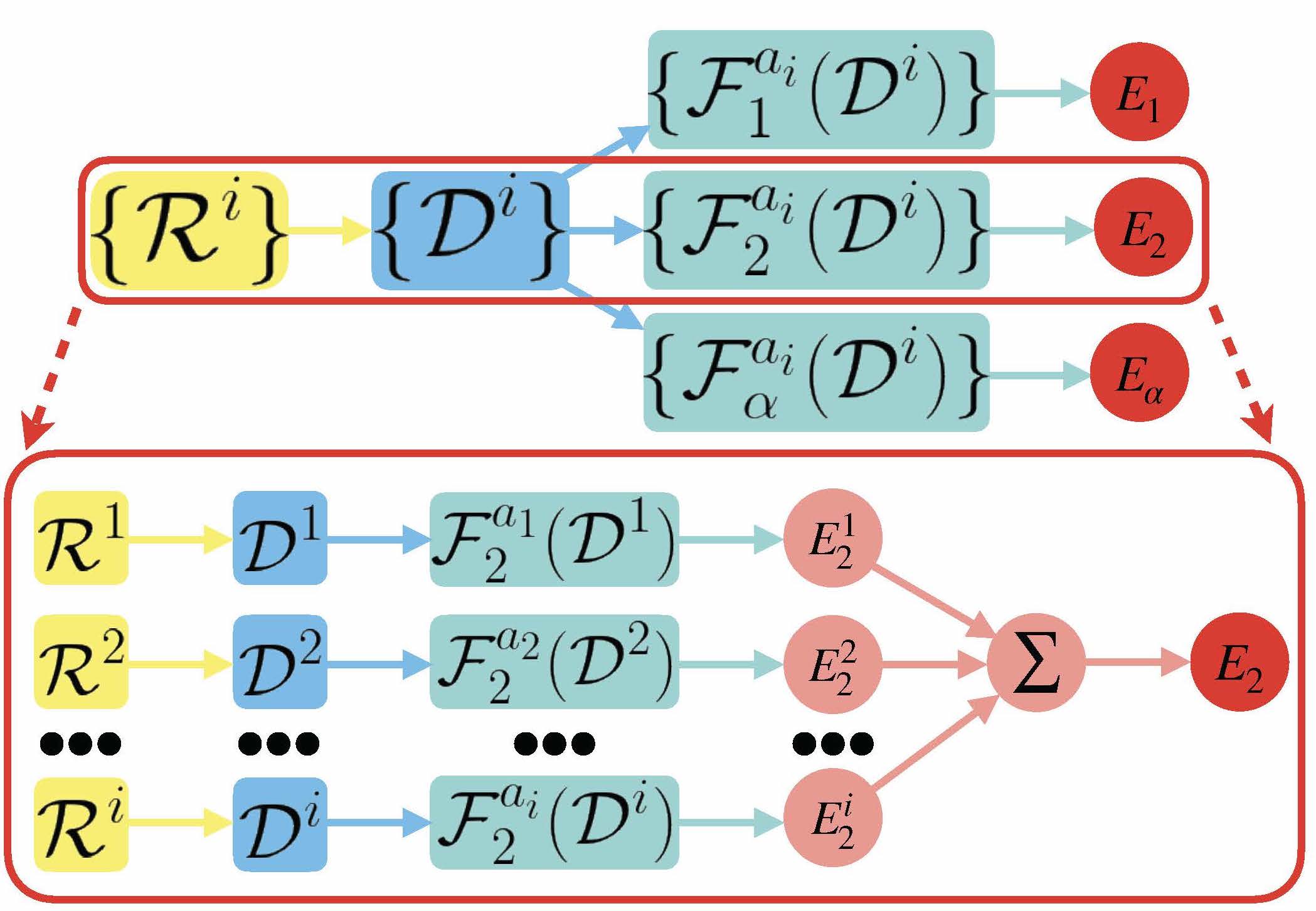}
    \caption{NN architecture for energy prediction. Coordinate matrices (yellow) are built for each atom local environment. They are then used to assemble descriptors (blue) with a 2-body embedding NN component. This collection of descriptor features is shared between all prediction components (green), each outputing the energy of a different state (red). A detailed view of the atom-wise contributions is shown for the first excited state. The NN architecture for NACV prediction within the CFA is similar in structure.}
    \label{fig:arch_FF}
\end{figure}

To compute the energies and forces of $n$ electronic states, extending the framework of DeePMD is straightforward: One can simply build a separate instance of the usual DeePMD architecture for each state considered, with descriptor and predictor components. 

In order to reduce computational complexity, we will use only one instance of the 2-body embedding descriptors of Eq. \eqref{eq:2bD}. This descriptor will be fed to all $n$ energy/force prediction components. Figure \ref{fig:arch_FF} schematizes the architecture used for energy prediction.
Parts of the NN dedicated to prediction of energy/forces are made completely independent from the components dedicated to predicting the NACVs (more precisely, $\mathbf{b}_{\alpha\beta}$). This is not a strict rule. It allows us to focus our attention on the learning of couplings, as learning the energy surfaces and forces is more routine. However, within both parts of the architecture, prediction of the different states/coupling will not be independent from the others as they will share embedding descriptors.
In addition to a decrease in complexity, an advantage of sharing descriptors between many prediction components is that it forces the embedding NNs to learn more general properties about the system than if one different embedding descriptor was generated per electronic state. Because it is put under greater constraint, the NN will be less prone to overfitting as a result.

For completeness, we mention a method for the prediction of vector properties which cannot be obtained as the gradient of a field, developed in DeepMD community\cite{deepmdkitv2} for the prediction of polarization \cite{zhang2020deep} through maximally localized Wannier functions.
After a fitting network $\mathcal{F}_1$ is fed with local atomic descriptor $\mathcal{D}^i$ given by Eq. \eqref{eq:2bD}, it outputs its last neuron layer of width $M$ which is combined with the coordinate matrix and embedding NN of Eq. \eqref{eq:2bG} in the following way:
\begin{equation}
\begin{split}
    \left( V_i \right)_{x} & = \frac{1}{N_c} \sum_{j=1}^{N_c} \sum_{m=1}^M \left( \mathcal{G}^i \right)_{jm} \left( \mathcal{R}^i \right)_{j,x+1} \left( \mathcal{F}_1 (\mathcal{D}^i) \right)_m \,   
    \\
    & x=1,2,3  
\end{split}
\end{equation}
A 3-dimensional full vector may be recovered by summing atom-wise contributions: $\displaystyle \mathbf{V} = \sum_i^{N} \mathbf{V}_i$. 

 Of present interest, we might consider whether this approach could be used to learn our NACV, however upon inspection, one realizes it would suffer from the problem described in the last paragraph of Sec.~\ref{sec:cfa}: 
Notice it effectively takes the vector space of all $\left( \mathcal{R}^i \right)_j$ as a basis to build vectors. 
These vectors are proportional to the cartesian position of atom $j$ in the frame whose origin is at the position of atom $i$ (see last 3 components of Eq.~\eqref{eq:GenR}), so when the molecule approaches a planar geometry,  all these vectors approach being parallel to  the plane. To represent a non-vanishing out-of-plane NACV component, some of the NN features in $\mathcal{G}^i$ and $\mathcal{F}_1 (\mathcal{D}^i)$ would need to spike to very high values when coming close to a planar geometry as a means to compensate the vanishing of out-of-plane components of all $\left( \mathcal{R}^i \right)_j$. The inherently smooth output of NNs will not be able to do so effectively, and the problem is reminiscent to the direct prediction of NACVs near CIs. 

\subsection{Predicting NACVs via their symmetric dyad}
\label{sec:learningdyad}
We now turn to our main contribution in this paper: to extend DeePMD to predict NACVs through learning the symmetric dyad matrix Eq.~(\ref{eq:Gamma}) described in Sec.~\ref{sec:symmdyad}. While Ref.~\cite{richardson2023machine} proposed the general approach, so far there has been no NN scheme to actually predict $\mathbf{\Gamma}$. Since the size of the matrix scales quadratically with the number of atoms and its elements can depend on up to two atom displacements, the question of how to build it efficiently from local descriptors is indeed nontrivial.  Our method is detailed below.

 As we do not have a conservative  field, we cannot rely on automatic differentiation to get vector components from a scalar field. Moreover, we cannot learn an effective vector from which to build $\Gamma$ as the multi-valuedness problem would reemerge. We thus need to predict elements of $\Gamma$ separately. The approach usually relied upon in DeePMD to build tensorial properties, such as the polarizability tensor\cite{sommers2020raman}, cannot be used for the reason mentioned at the end of Sec.~\ref{sec:property}  concerning vector prediction.
Hence, we devised a new approach exploiting symmetries of $\Gamma$  and the atomic embedding principle of DeePMD.

We learn  $\Gamma_{\alpha\beta}$ for each pair of states $\alpha,\beta$ as a collection of 3 by 3 blocks 
\begin{equation}
\Gamma_{\alpha\beta} =\begin{bmatrix}
    \Gamma^{11} & \Gamma^{12} & \Gamma^{13} & . \\
    \Gamma^{21} & \Gamma^{22} & \Gamma^{23} & . \\
    . & . & . & .
\end{bmatrix}_{\alpha\beta}
\end{equation}
where each block has the structure
\begin{equation}
\Gamma_{\alpha\beta}^{ij} =\begin{bmatrix}
    \Gamma^{ij}_{xx} & \Gamma^{ij}_{xy} & \Gamma^{ij}_{xz}  \\
    \Gamma^{ij}_{yx} & \Gamma^{ij}_{yy} & \Gamma^{ij}_{yz} \\
    \Gamma^{ij}_{zx} & \Gamma^{ij}_{zy} & \Gamma^{ij}_{zz}
\end{bmatrix}_{\alpha\beta}
\end{equation}
Blocks on the diagonal $i=j$ are obtained from components of the NACV associated to a single atom, and thus the descriptor $\mathcal{D}^i$ associated to it will suffice. It is fed to a fitting neural network set to output the diagonal block. More precisely, only the upper triangular block needs to be predicted, the lower part will be obtained by symmetry: 
\begin{equation} \label{eq:dGP}
    \Gamma_{\alpha\beta}^{ii} = \mathcal{F}_{\alpha\beta}^{ii}\left( \mathcal{D}^i \right)\,.
\end{equation}
The off-diagonal blocks contain information related to two atoms, so the descriptors $\mathcal{D}^i$ and $\mathcal{D}^j$ associated to both relevant atoms should be fed to the fitting neural networks.  We take the simple strategy to hybridize the two descriptors and feed them to the prediction NN. All elements of off-diagonal blocks are different in principle, and they will be obtained explicitly from the fitting network:
\begin{equation} \label{eq:GP}
    \Gamma_{\alpha\beta}^{ij} = \mathcal{F}_{\alpha\beta}^{ij}\left( (\mathcal{D}^i, \mathcal{D}^j) \right)
\end{equation}
To reduce the complexity and enforce permutation symmetry, NN parameters of diagonal blocks $\mathcal{F}_{\alpha\beta}^{ii}$ will be shared for all atoms of same species, and similarly for all pair of species in off-diagonal block predictors $\mathcal{F}_{\alpha\beta}^{ij}$. 

 We then have to account for the effect of overall rotation of the system on blocks of $\mathbf{\Gamma}$. To this end, we use rotation matrices $\Omega^{i}$ linked to local frames of each atom.  The specific choice of definition for the local frames does not matter as long as it evolves continuously with deformation of the molecule.  For our test example of CH$_2$NH$_2^+$, we simply define the $z$-axis along the CN bond for all $\Omega^{i}$, while the $x$-axis depends on atom $i$. If it is a hydrogen, the $xz$-plane contains the H atom and the CN bond. Orthogonality allows to deduce where the $x$ and $y$ axes lie. For C and N atoms, { the first hydrogen atom} was picked to define the $xz$ plane as explained above. { Graphs illustrating the procedure are provided in Sec. I. of Supplementary Material.} 

Each predicted block of $\Gamma_{\alpha\beta}$ obtained from Eqs. \eqref{eq:dGP} and \eqref{eq:GP} is then rotated 
according to
\begin{equation}
    \Gamma_{\alpha\beta}^{ij} \longrightarrow \Omega^{i} \; \Gamma_{\alpha\beta}^{ij} \; (\Omega^{j})^T
\end{equation}

{ We can extract the desired $\bb_{\alpha\beta}$ from the eigenvector corresponding to the non-zero eigenvalue (see Sec.~\ref{sec:symmdyad})}. 
{ As the dyad has only one non-zero eigenvalue, it is not necessary to perform an explicit diagonalization to obtain $\mathbf{b}_{\alpha \beta}$: it can be obtained from any column by simply dividing it with the square root of the corresponding diagonal element. Doing so allows to lower the computational cost. We will use this procedure throughout this work and always choose the column corresponding to the highest diagonal element in order to avoid potential instabilities.}

The last step in building our NN architecture is to provide the descriptors $\mathcal{D}^i$. 
 Numerical tests show existing DeePMD descriptors lead to important errors on $\Gamma_{\alpha\beta}^{ij}$. Two-body embedding incorrectly predicts same values of NACV components for { some} geometries differing by a rigid torsional motion around the C-N bond { (see Sec. S-II of Supplementary Material for a detailed analysis)}.  It is important to underline that this shortcoming is the consequence of the different way we use atom-wise descriptors to predict couplings: We do not sum atom-wise contributions as when predicting energies, instead each atom-wise descriptor should hold sufficient information to predict a (diagonal) block of $\Gamma_{\alpha\beta}$. The three-body embedding descriptor does hold more angular information and  using Eq.~\eqref{eq:3bD} did result in a clear improvement, but not to a completely satisfactory level. We observe that the angular information is essentially 'summed over', which was not a problem in the context of force prediction through automatic differentiation  since the gradient of this descriptor would  contain this information, but our approach avoids this since our field is not conservative.

To remedy this, we introduce a new descriptor built on the same elements as the 3-body descriptor described above. We first feed the angle information to an embedding neural network to obtain $\mathcal{G}^i = \mathcal{N}_{e,3}\left( \theta^i \right) \in \mathbb{R}^{N_c \times N_c \times M}$. Then, to each matrix $N_c \times N_c$ matrix contained in $\mathcal{G}^i$, we apply successive matrix multiplications with the last three columns of $\mathcal{R}^i$ from the right and its transpose from the left. This defines the  intermediate
descriptor $\mathcal{D}^i \in \mathbb{R}^{M \times 3 \times 3}$:
\begin{equation} \label{eq:GD}
    \left(D^i\right)_{m} = \sum_{k=1}^{N_c} \sum_{l=1}^{N_c}  \Big(\left( \tilde{\mathcal{R}}^i \right)_k\Big)^T 
    \left(\mathcal{G}^i\right)_{klm} \; \left( \tilde{\mathcal{R}}^i \right)_{l}
\end{equation}
\begin{equation} \label{eq:Rt}
    \left( \tilde{\mathcal{R}}^i \right)_k = \left\{  \frac{s(r_{ik})x_{ik}}{r_{ik}} \quad \frac{s(r_{ik})y_{ik}}{r_{ik}} \quad \frac{s(r_{ik})z_{ik}}{r_{ik}} \right\}
\end{equation}
 Notice that, unlike the three-body descriptor of Eq.~(\ref{eq:3bD}), this descriptor does not contract angular information. The fact that angular information is not summed over would ease the job of the prediction layer, but it also means the descriptor is not rotationally invariant. We need to enforce invariance before feeding it to a fitting network. To this end, we again make use of rotations matrices introduced above:
\begin{equation}
    \mathcal{D}^i  =\sum_{k=1}^{N_c} \sum_{l=1}^{N_c}  \Big(\left( (\Omega^{i})^T \tilde{\mathcal{R}}^i \right)_k\Big)^T 
    \left(\mathcal{G}^i\right)_{klm} \; \left( (\Omega^{i})^T \tilde{\mathcal{R}}^i \right)_{l}
\end{equation} 
Our method overcomes all above-mentioned limitations of the CFA and previous DeePMD strategies while being numerically efficient by sticking to the atomic embedding principle. 

\begin{figure}[h!]
    \centering
\includegraphics[width=.5\textwidth]{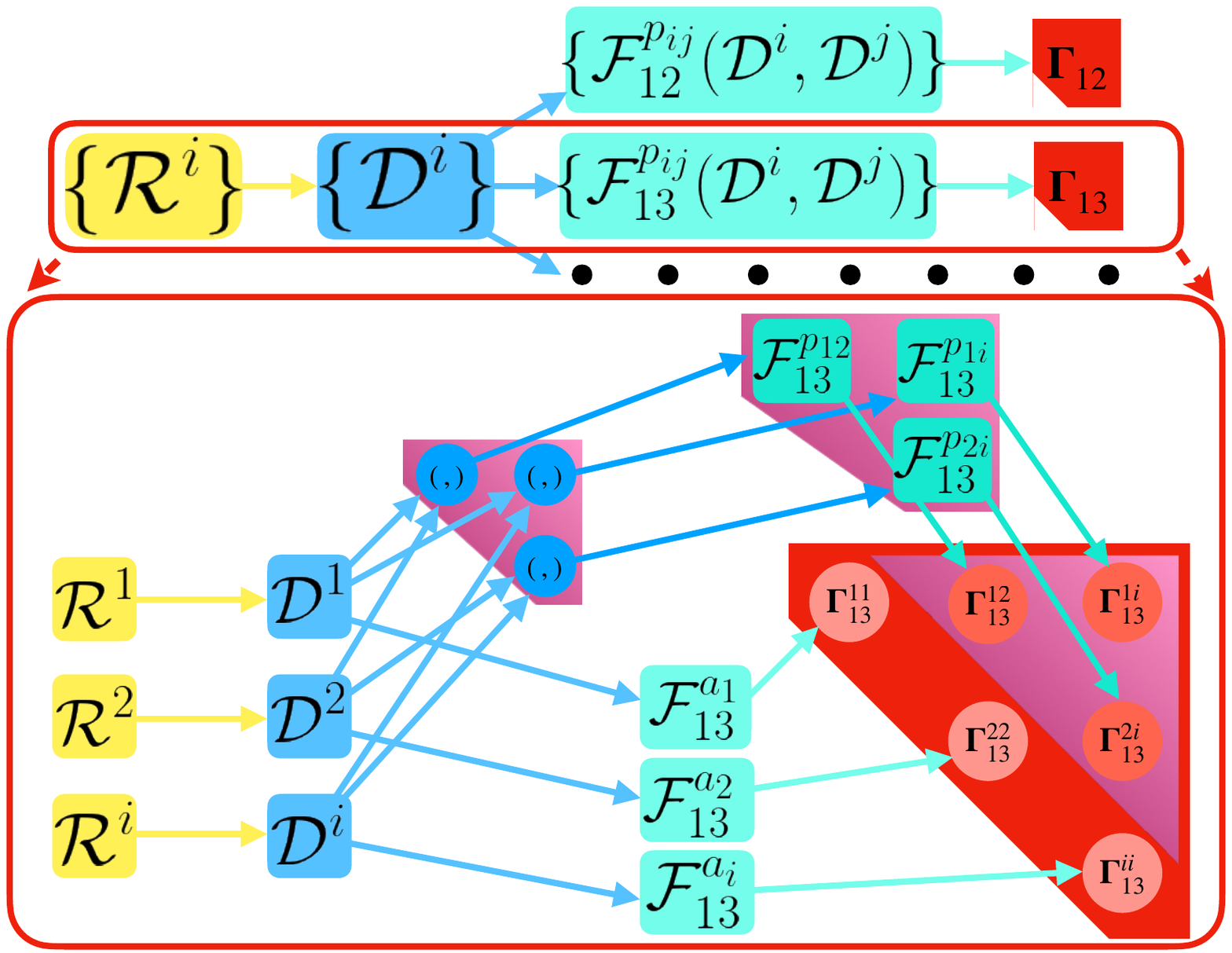}
    \caption{NN architecture for dyad prediction. Coordinate matrices (yellow) are used to construct our generalized descriptors (blue) with a 3-body embedding NN component. This collection of descriptor features is shared between all prediction components (cyan) outputing the (upper triangular parts of) dyads of all required couplings. Atom-wise contributions within the components are shown for the coupling between states 1 and 3. Detail of the prediction of diagonal and off-diagonal blocks of the dyad is sketched.}
    \label{fig:arch_G}
\end{figure}

In our application of the dyad matrix method,  the architecture thus separates prediction of energies/forces from prediction of (the numerator of the) NACVs.

 To predict the couplings  between the $n$ states, $n(n-1)/2$ different $\mathbf{\Gamma}_{\alpha\beta}$ matrices need to be built. To do so, a specific prediction component per coupling will be built, with 3-by-3 sub-blocks of $\mathbf{\Gamma}$ predicted using fitting NNs defined in Eq. \eqref{eq:GP}. Following the same principle as before, all those prediction components will be fed with the same single instance of descriptor given by Eq. \eqref{eq:GD}. Figure \ref{fig:arch_G} sketches the architecture for dyad prediction.

 For clarity and conciseness, we presented the architectures omitting the fact that NN parameters are shared between atoms of same species $a_i$ or pair of species $p_{ij}$. If we view two NNs of the same structure but with different weights and biases as two separate entities, we remind the reader that there will be as many $\mathcal{N}_{e,3}^{p_{jk}}$ as there are different pairs of atomic species, and similarly for $\mathcal{F}_{\alpha\beta}^{jk}$. Skip connection\cite{he2016deep} with randomized weights lying within the range $0.1 \, \pm \, 0.001$ was used in all NNs.

\section{Training method}
\label{sec:training}
 We now turn to the methods used for training the NNs.  We developed an in-house python code using JAX\cite{jax2018github}, FLAX\cite{flax2020github}, and OPTAX\cite{deepmind2020jax} to define the NNs and setup the training.   

\subsection{Loss functions}
We train the NN using Adam stochastic gradient descent method\cite{kingma2014adam} combined with weight decay~\cite{loshchilov2017decoupled}.

The loss function comprises  L2 losses associated to the energies, forces and NACVs:
\begin{equation}
    L_{\text{D}} = L_{E} +  L_{\mathbf{F}} + L_{\mathbf{\Gamma}} \,.
\end{equation}
For the energies we have
\begin{equation}
    L_{E} = \frac{1}{N_b}\displaystyle \sum_l^{N_b} \sum_{\alpha}^n \Big| E_{l,\alpha}^{\text{Ref}} - E_{l,\alpha}^{\text{ML}} \Big|^2
\end{equation}
where $N_b$ is the total number of elements in the training batch. Similarly, for the forces we define
\begin{equation}
    L_{\mathbf{F}} = \frac{1}{N_b}\displaystyle \sum_l^{N_b} \sum_{\alpha}^n \Big|\Big| \mathbf{F}_{l,\alpha}^{\text{Ref}} - \mathbf{F}_{l,\alpha}^{\text{ML}} \Big|\Big|^2\,,
\end{equation}
 where $|| \mathbf{v} ||^2$ denotes the averaged square of all components of the vector $\mathbf{v}$,
and for the symmetric dyad,
\begin{equation}
    L_{\mathbf{\Gamma}} = \frac{1}{N_b}\displaystyle \sum_l^{N_b} \sum_{\alpha<\beta}^n \Big|\Big| \mathbf{\Gamma}_{l,\alpha\beta}^{\text{Ref}} - \mathbf{\Gamma}_{l,\alpha\beta}^{\text{ML}} \Big|\Big|^2\,.
\end{equation}

\subsection{Input and output normalization}
Input and output normalization are used to facilitate the search for the optimal learning rate and control the range spanned by features throughout the neuron layers. Letting $x$ represent the coordinates or energies  in the training set of size $N_s$, we use its mean $\Bar{x}$ and standard deviation $\sigma_x$,
\begin{equation}
    \Bar{x} = \frac{1}{N_s} \sum_{l}^{N_s} x_{l} \;\;\;{\rm and}\;\;\; \sigma_x = \sqrt{\frac{1}{N_s}\sum_{l}^{N_s} \Big( x_{l} - \Bar{x} \Big)^2}\,.
\end{equation}
to normalise the values of the switching function $s(x)$ in the following way:
\begin{equation}
    \tilde{x}_l = \frac{x_l-\Bar{x}}{\sigma_x}
\end{equation}
 The normalized coordinates are fed into the NN while the normalized reference energies are fed into the loss function.
This implies that the forces are to be normalized as well following
\begin{equation}
    \tilde{\mathbf{F}}_{\alpha,l} = \frac{\mathbf{F}_{\alpha,l}}{\sigma_{E_\alpha}}.
\end{equation}
Energies and forces corresponding to a given electronic state $\alpha$ are normalized independently from other states.

Concerning the NACvs, $\mathbf{\Gamma}$ matrices in the loss function are normalized following:
\begin{equation}
   \tilde{\mathbf{\Gamma}}_{\alpha\beta,l} = \frac{\mathbf{\Gamma}_{\alpha\beta,l}}{\displaystyle\sqrt{\frac{1}{N_s}\sum_{l}^{N_s} \mathbf{\Gamma}_{\alpha\beta,l}^2}}
\end{equation}
Each coupling between a pair of states is normalized separately over the dataset.
After training, then the NN will be used to run dynamics, its outputs will be rescaled inversely. 

\subsection{Hyperparameter optimization}
\label{sec:hyperparameter}
Hyperparameter optimization will rely on a training dataset and a smaller validation dataset that will be described in next section.
We start the optimization of hyperparameters with the determination of an optimal learning rate.  During this phase, only the performance on the training set is monitored. To do so, a relatively small ML structure is used to speed-up the numerous training processes that are performed at this stage. To learn energies and forces, embedding/fitting NNs are built with 3 neuron layers of width $M=32$ and a value $M_<=12$ for the 2-body embedding descriptor (see Eq. \eqref{eq:2bD}). 
To learn $\mathbf{\Gamma}$ matrices, separate 3-body descriptors are built from their own embedding NN composed of 3 neuron layers of width $M=32$. They are then fed to fitting networks with 3 neuron layers of width $M=32$.

Initially, the learning rate is sampled uniformly (in logscale) in the range $[0.01,0.00001]$ and training is stopped early, at 625 epochs. The batch size was set to 50. After 3 cycles of refining the grid for the sampling of the rate in the range of optimal performances, the best performing learning rate $\gamma^{(0)}_{\rm best}=0.0012$ is kept. We follow-up by increasing the number of training epochs to $2500
$,  set the initial learning rate to $\gamma^{(0)}_{\rm best}$ and sample different values of final learning rate uniformly in the range $[0.0012,0.00001]$.  The transition from initial to final learning rate is done by exponential decay, the learning rate being changed every 10 descent steps. The optimized value was $\gamma^{(f)}_{\rm best}=0.00005$. We keep the best performing values of initial/final learning rate going forward. 

We then increase the width $M$ of each neuron layer until a balance is struck between maximizing accuracy and preventing overfitting. This is done by checking the ability of the NN to learn the training dataset while performing well on the validation set. In the present study, we consider the reproduction of population dynamics obtained without ML as the true 'test' of our NN (see Results), as the ab initio trajectories used as reference are not part of the datasets used.
A value of $M=40$ for all NN widths together with a weight decay coefficient of $0.005$ allowed a satisfying performance. We kept $M_<=12$ to minimize computational cost.

Owing to the molecular scale of the systems considered in the present work, optimization of the  number of atoms included in the local environment descriptors is superfluous at the start of the hyperparameter search. We thus started with cutoff radii of $r_s=11.8$ a.u. and $r_c=12$ a.u. so that all atoms are included in their respective neighborhood  in previous trainings. Then, cutoff were decreased to the minimum values which did not compromise accuracy. Those values turned out to be $r_s=5$ a.u. and $r_c=7$ a.u.

\section{Results: the methaniminium cation CH$_2$NH$_2^+$}
\label{sec:results}
 We use the methaniminium cation to test our implementation, paying particular attention to the prediction of the NACVs. This system was used in the work of Ref.~\cite{westermayr2019machine,westermayr2020combining} to test their implementation of machine-learned energies, forces, and NACVs in the { SchNet} code. We use the data provided in { Supplementary Material of this previous study}~\cite{westermayr2019machine}, comprised of a training set of 4000 datapoints and a validation set of 770 datapoints, obtained using the MR-CISD(6,4)/aug-cc-pVDZ electronic structure method. Those data were phase-corrected by the original authors, but we note that our dyad method is impervious to any phase choices.

\subsection{Performances on the dataset}

We first compare the capability of the CFA and the dyad method to learn the NACV orientation. We refer to the molecular plane at equilibrium geometry as the { $xz$-plane} in the following.

\begin{figure}[!htbp]
    \centering
    \includegraphics[width=.4\textwidth]{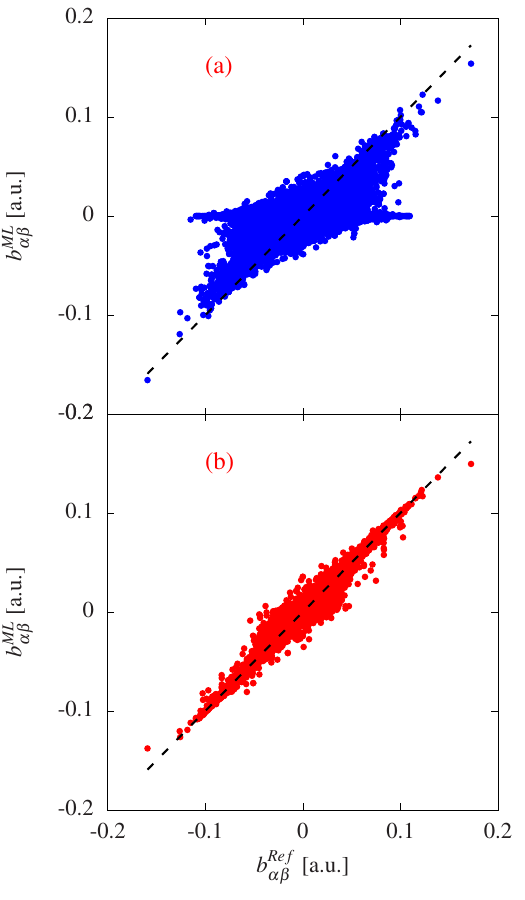}
    \caption{Performance of ML methods to reproduce out-of-plane $b_{\alpha\beta}$ components on the training set. Learned values are plotted against reference values. Panel (a): CFA. Panel (b): dyad method.}
    \label{fig:bvec_y_cf_vs_dyad}
\end{figure}

In Figure \ref{fig:bvec_y_cf_vs_dyad}, machine-learned values of the $y$-components of $\bb_{\alpha\beta}$ are plotted against reference values over the whole dataset. 
Panel (a) and (b) are obtained using CFA and dyad method, respectively. The CFA is seen to prevent the NN to learn the out-of-plane component, even to a qualitative level. In addition to the numerous values predicted as essentially zero for near-to-planar geometries, the whole set is learned with a much lesser accuracy compared to the dyad method, as illustrated in Table~\ref{tab:ML_vs_QC} where the mean absolute error (MAE) and root mean square error (RMSE) are given for each method, separating each cartesian component of $b_{\alpha\beta}$.
This shows that the underlying assumption in CFA of zero curl of $\bb_{\alpha\beta}$ has a significant error in its accuracy. The dyad method has no such inherent approximation and the machine-learned values are significantly more accurate, particularly along the molecular plane of symmetry.

\begin{table}[!htbp]
\centering
\caption{Mean absolute error (MAE) and root mean square error (RMSE) on $b_{\alpha\beta}$ over the training set using dyad method and CFA for each cartesian component. }
\begin{tabular}{p{1.5cm} | p{1.5cm} p{1.5cm} p{1.5cm}}
\hline
\; Method &  \multicolumn{3}{c}{MAE (RMSE) [ $10^{-3}$ a.u.]} \\
\hline
{}   & \quad \quad $x$   & \quad \quad $y$    & \quad \quad $z$   \\
\quad \; CF & \quad 5.4 (8.2) & \quad 16 (24) & \quad 8.0 (14)  \\
       \quad Dyad & \quad 1.9 (3.2) & \quad 2.1 (3.8) & \quad 1.9 (3.1)
       \\ \hline
\end{tabular}    \label{tab:ML_vs_QC}
\end{table}

{ One could wonder how much of the improvement is due to the new descriptor we introduce rather than to the dyad method in itself. In Sec. S-II of Supplementary Material, we show that combining our more complex descriptor with CFA predictor does not improve significantly the accuracy of the CFA. The wealth of information it provides is only needed and taken advantage of by the dyad method for which it was specifically designed.}

{ We also analyzed the quality of NACV prediction by the CFA and dyad methods at datapoints lying in the vicinity of CIs. Details about the datapoints chosen and the effective coordinates they scan are given in Sec. S-III of Supplementary Material. In Figure~\ref{fig:CI_y_cf_vs_dyad} we compare the performance of ML predictions for both methods. The left panels show results in the S1/S0 CI vicinity and right panels show results in the S2/S1 CI vicinity. The top panels show the predicted  norms of the NACV predicted by CFA (blue squares) and dyad (red circles) compared to the reference values (grey line). It should be noted that in an effort to isolate the error of the predicted $\textbf{b}_{\alpha\beta}$ from that of the predicted energy gaps appearing in the
denominator of the NACV, in these figures we instead divide the predicted numerator $\textbf{b}_{\alpha\beta}$ by the exact energy gaps. It is seen that for both CIs, the dyad methods reproduces the NACV norm very faithfully while the CFA approach fails to do so. The middle panels show how well the direction of the NACV is reproduced by ML methods. The collinearity is a normalized dot-product of the machine-learned NACV with the reference NACV, defined in Sec. S-III of the Supplementary Material, and reaches 1 when the direction is perfectly matched while decreasing to 0 if the prediction is orthogonal to the reference vector. We see that the dyad matches the reference NACV values in the vicinity of CIs both in magnitude and in direction. Note that using the ML-predicted energy gaps to compute NACVs, as is done in practice, is the biggest source of error for dyad prediction in this work; see figures in Sec. S-III of Supplementary Material. However, even with approximate energy gaps, these figures show that the dyad method is a substantial improvement over the CFA approach.  } { We also noticed a difference in performance of the CFA method for the S$_0$/S$_1$ CI vicinity depending on the loss function used for training: Results using the phaseless loss shown in Fig.~\ref{fig:CI_y_cf_vs_dyad} are noticeably better than with the L2 loss (see Fig. S9 in Supplementary Material).}

\begin{figure}[!htbp]
    \centering
    \includegraphics[width=.5\textwidth]{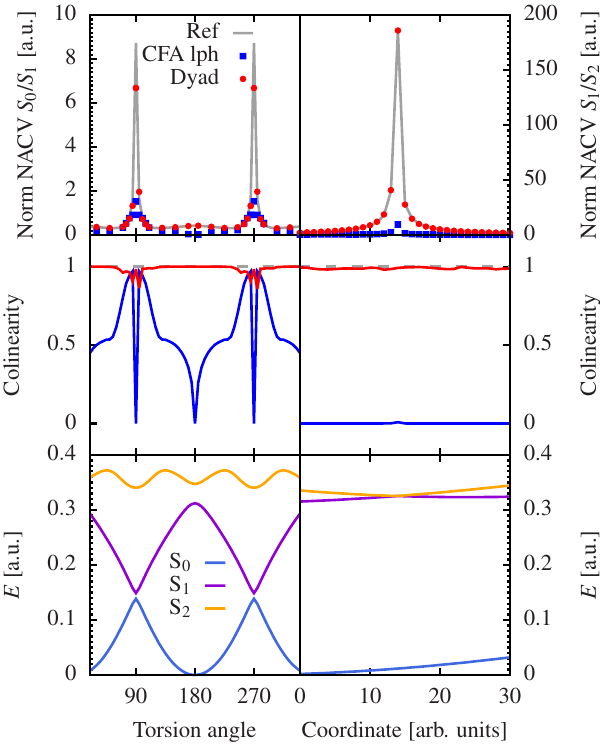}
    \caption{Performance of ML methods to reproduce NACV norm and orientation at datapoints in the vicinity of both CIs. CFA was trained using phaseless loss (lph). Left and right panels show results around the S$_1$/S$_0$ and S$_2$/S$_1$ CIs respectively. 
    Top panels: Norm of NACVs obtained by scaling ML predictions of $\mathbf{b}_{\alpha\beta}$ with exact energy gaps compared to reference values. 
    Middle pannels: colinearity of predicted NACVs with reference NACVs (see Supplementary Material for exact formula). 
    Bottom panels: Exact energies.}
    \label{fig:CI_y_cf_vs_dyad}
\end{figure}

 We stress that predicting the NACV magnitude correctly is not enough to ensure the dynamics will be accurate. Reliably predicting the direction of NACVs is essential  since electronic population transfer is mediated by the projection of the NACV along the trajectory's velocity. Further, in general trajectory-based methods (although not surface-hopping) the direction of the NACV directly influences the force on the nuclei.
Within the context of surface hopping simulations, not only the hopping probability is strongly affected by the relative direction of the trajectory's velocity and the NACV, but also the 
velocity adjustment procedure is typically dependent on this:
The most widely used ansatz to perform velocity adjustment after a hop is to change the velocity component collinear to the NACV\cite{carof2017detailed,barbatti2021velocity,tang2021evaluation,vindel2021study,huang2023first,dupuy2024exact}, so if this NACV direction is wrongly predicted, the trajectory's path after a hop will be erroneously altered. For example, if the NACV is wrongly predicted to lie along a direction where the trajectory has low momentum, not only the hopping probability would be underestimated, but it could 
also cause a spurious frustrated hop to occur. 
Further, if the velocity-reversal method was chosen to treat frustrated hops~\cite{sifain2016communication}, this would risk
erroneous reflections of trajectories at every time step. 

\subsection{SHEDC dynamics}
We performed surface-hopping with energy-based decoherence correction (SHEDC) dynamics~\cite{T90,granucci2007critical,granucci2010including} using our NN with dyad method to supply energies, forces and NACVs. Details regarding initial conditions and dynamical parameters are identical to that of Ref~\cite{westermayr2019machine}, and recalled below. We used a total of 1000 trajectories, for which initial conditions were obtained via Wigner sampling along normal modes taken at the ground state equilibrium geometry. We checked  that at all  initial conditions, the machine-learned energy gap between $S_0$ and $S_2$ lay within $9.44 \pm 0.15$ eV. They were then propagated for 100 fs, starting in $S_2$.
The integration timestep was set to 0.5 fs for nuclear motion, with 25 electronic timesteps performed within each nuclear timestep. As NACVs  are defined up to an arbitrary sign, phase tracking was performed along the trajectories. Velocity rescaling is performed along NACVs, and frustrated hops result in reversing the velocity. 

\begin{figure}[!htbp]
\centering
\begin{subfigure}{0.5\textwidth}
\includegraphics[width=\textwidth]{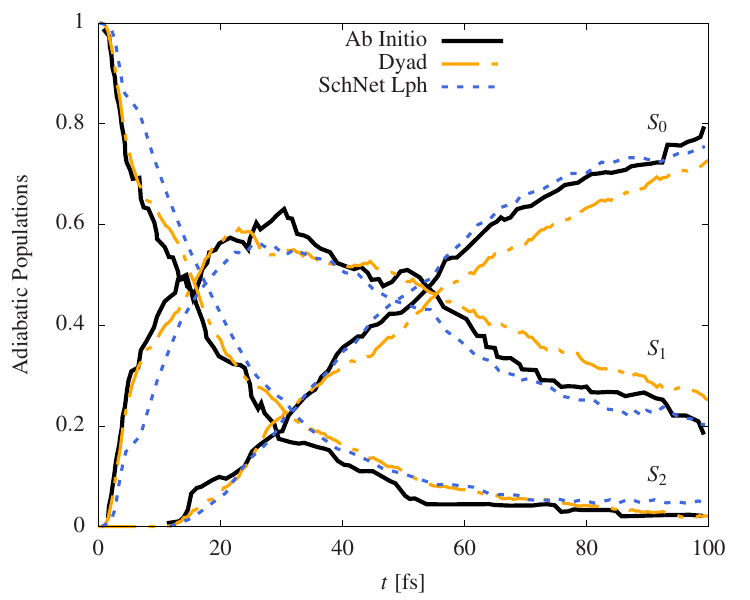}    \caption{}
    \label{fig:first}
\end{subfigure}
\hfill
\begin{subfigure}{0.5\textwidth}
\includegraphics[width=\textwidth]{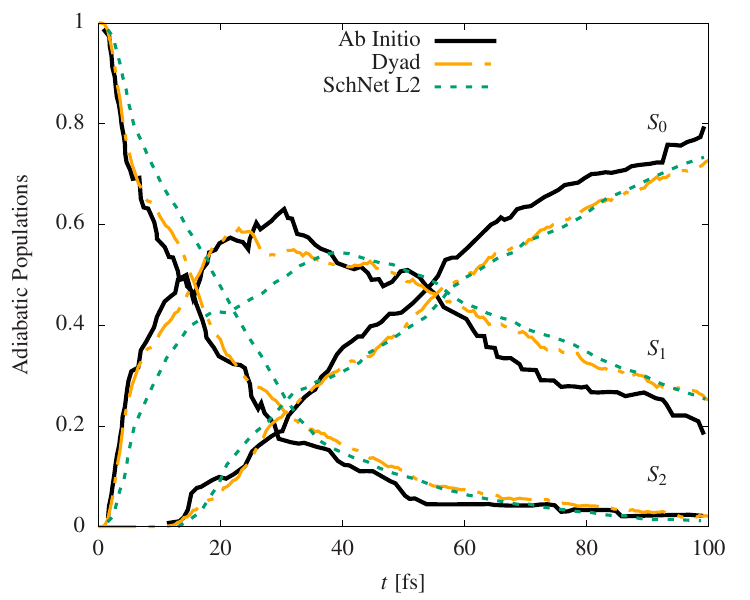}    \caption{}
    \label{fig:second}
\end{subfigure}
\caption{Comparison of population dynamics relying on explicit MR-CISD(6,4) calculations (ab initio, see Ref \cite{westermayr2019machine}) in black solid lines to our results using the dyad method in orange dashed lines. On panel (a), results of Ref \cite{westermayr2020combining} using the CFA and phaseless loss in training is shown in blue dashed lines. On panel (b), results using L2 loss are shown in green dashed lines. All NNs were trained on a phase-corrected dataset. }
\label{fig:dyn_d_cf_pc}
\end{figure}

In Figure \ref{fig:dyn_d_cf_pc} we compare the resulting population dynamics (dashed orange lines) to reference results of 90 trajectories propagated from the MR-CISD(6,4)/aug-cc-pVDZ electronic structure method (solid black lines). The agreement is very satisfactory, and is compared to previous results of Ref \cite{westermayr2020combining} using ML relying on the CFA to replace the {\it ab initio} calculations. All ML results shown in Figure~\ref{fig:dyn_d_cf_pc} exploit NNs that were trained on the same dataset, comprised of 4000 datapoints that were phase-corrected before training\cite{westermayr2019machine}. All population dynamics obtained through ML employ the same number of trajectories. Panel (a) shows results using CFA together with the phase-less loss Eq.~\eqref{eq:lph} during training of the NN in blue dashed lines. The green-dashed lines in panel (b)  also shows the CFA but using instead the L2 loss, 
$L_{\text{NAC}}=  \frac{1}{N_b}\displaystyle \sum_l^{N_b} \sum_{\alpha<\beta}^n \Big|\Big| \mathbf{d}_{l,\alpha\beta}^{\text{Ref}} - \mathbf{d}_{l,\alpha\beta}^{\text{ML}} \Big|\Big|^2 $  during training. As the dataset is pre-processed through a phase correction scheme, use of the phase-less loss function rather than L2 loss should yield the same accuracy if this pre-processing is robust. However, the NN trained using the L2 loss is seen to predict slightly slower population transfer from $S_2$ to $S_1$ during the first 20 fs and from $S_1$ to $S_0$ at later times. We attribute these variations to the non-rigorous character of the dataset phase pre-processing, aiming at making a single-valued, smooth NACV field out of a fundamentally multi-valued one\cite{richardson2023machine}. 
Moreover, both CFA-based results exhibit slower population transfer from $S_2$ to $S_1$ at early times compared to ab initio results. Our dyad method yields population dynamics in excellent agreement with ab initio during the first 50 fs, followed by a slightly slower population transfer between $S_1$ and $S_0$ at later times. Owing to the limited size of the training dataset, together with the fact it was generated from adaptive sampling relying on the CFA approximation\cite{westermayr2019machine}, it is possible our dyad trajectories explore a region of configuration space that was not sampled as comprehensively. Moreover, as only 90 ab initio trajectories were propagated, it is not certain the ab initio populations dynamics do not suffer from undersampling at later times. We are thus led to consider our quantitative prediction for early times as a significant proof of the robustness of our dyad method.

\begin{figure}[!htbp]
    \centering
    \includegraphics[width=.5\textwidth]{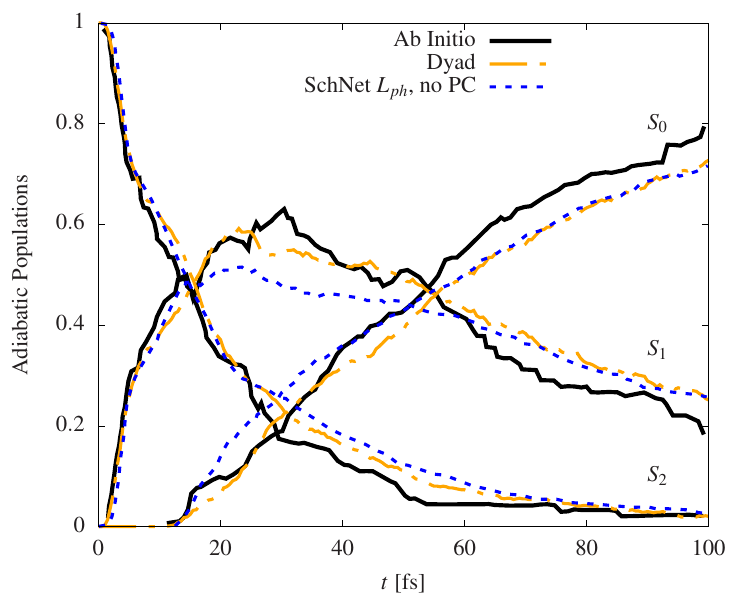}
    \caption{Comparison of population dynamics relying on explicit Quantum Calculations (QC, see ref \cite{westermayr2019machine}) in black solid lines to our results using the ML dyad method in orange dashed lines. Results of ref \cite{westermayr2020combining} using ML with CFA trained on data without any phase correction applied before hand (the phaseless loss is thus used during training) are shown in blue dashed lines. }
    \label{fig:dyn_d_cf_nopc}
\end{figure}

Lastly, we compare our results using the dyad method (orange dashed lines) to the CFA NN of ref \cite{westermayr2020combining} trained on not-phase-corrected data using the phase-less loss (blue dashed lines) on Figure \ref{fig:dyn_d_cf_nopc}. The fact that  the latter results in different dynamics is another illustration of the uncontrolled effect of approximating the NACV as if it was a conservative field. Even if our dyad NN was trained on phase-corrected data, as our method circumvents the multi-valuedness and sign-arbitrariness problems, it is impervious to the phase correction pre-processing. This justifies our choice to compare its performance to CFA NNs no matter how they were trained. Again, the rigorousness of our dyad approach results in its better and more robust performance. 

Still, for this system, the CFA approximation is seen to perform reasonably well, especially when training on not-phase-corrected data. The moderate impact on the dynamics of the absence of out-of plane components of NACVs in CFA NNs can be explained by the fact that for this system, the torsional motion around the C-N bond that would be expected to mediate population transfer is not the main relaxation path, as already observed in a previous study \cite{barbatti2006ultrafast}. Trajectories acquire a large momentum leading to important stretching of the C-N bond when they are initialized on the $S_2$ state, allowing to reach the intersection seam between $S_0$ and $S_1$ in combination with a bi-pyramidalization motion. We expect our dyad method will exhibit a bigger impact on the quality of dynamics for other systems.

\section{Conclusions and Outlook}
\label{sec:conc}
In summary, we extended the DeePMD NN architecture to predict excited state energies, forces, and couplings necessary to perform non-adiabatic dynamics simulations. We introduced a method to efficiently learn the symmetric dyad of non-adiabatic coupling vectors NACVs, overcoming several issues of the conservative field approximation (CFA) used until now in other NN architectures. 
The lack of Berry phase effects in the CFA approach has been recognized in the literature~\cite{westermayr2019machine,westermayr2020combining,WM20rev}, and downplayed because the NN are typically used in conjunction with mixed quantum-classical methods which are unable to capture those effects in the first place.
 However, the impact of neglecting the curl of { the energy-difference scaled} NACVs goes beyond this effect and represents an uncontrolled approximation even in cases without CIs. Here, we showed that    learning the energy-difference scaled NACVs as we do force fields strongly impedes the capability of ML to learn them faithfully because of its complete failure to reproduce components orthogonal to a plane of symmetry.  We believe this point should be brought to greater attention, as this major source of error is separate from considerations of accumulated phases around CIs. Both a wrong magnitude and wrong direction of the NACV can influence the dynamics no matter what mixed quantum-classical method is used, and can obscure interpretation of which molecular motion funnels population transfer.
 Application to the photodynamics of the methaniminium cation showed our approach achieves quantitative accuracy, and its rigorous basis constitutes a more robust approach for ML-powered  non-adiabatic dynamics. The better performance  in early time dynamics is much more significant
than the (still small) disagreement with reference calculation in long-time dynamics once we note that the latter used
only 90 trajectories: While this number is sufficient to sample stochastic hops between electronic states for a short time, SH simulations typically need a few hundred trajectories to reach statistical convergence over the full relaxation process. Moreover, the dataset used to train our NNs is biased towards the
CFA dynamics as it was built through adaptive sampling along CFA trajectories in Ref.\cite{westermayr2019machine}, which could lead to undersampling of
pockets of configuration space relevant to long time dynamics if dyad-method
trajectories end up evolving in different regions. Although at earlier times, the CFA-
method results are worse than the dyad method, they are still in good agreement with
reference calculations, and this is largely due to the relaxation dynamics in this case
largely avoiding torsional motion.

{ Considering the computational scaling of the dyad method with system size, although the dyad matrix grows quadratically with the number of atoms, the number of NN parameters used only grows with the number of different atomic species. Moreover, we do not need to perform explicit diagonalization of the dyad, because NACVs can be extracted from a single column. In the future, this property could be exploited even more by only generating a very small subset of the matrix at the prediction stage: predict the diagonal to identify the biggest element, then predict the associated column. This will help the approach to scale favorably with system size.}

 {  Our analysis focused on DeepMD and SchNet descriptors for which we discussed how the CFA approach results in strict molecular symmetry constraints. Many other approaches to learn force fields exist that were not discussed here. It is however understood that they all strive to exploit molecular symmetries in order achieve better generality and rigorousness, would it be by the use of symmetry functions as in the Behler-Parinello approach\cite{behler2007generalized} or O(3)-equivariant graph neural networks in MACE\cite{batatia2022mace}. As a consequence, directly applying force-field-specific ML techniques to NACV prediction would be ill-advised, but rather new adaptations of these schemes should be developed along the lines of this work. In addition to symmetry considerations, and the neglect of any rotational part of the energy-difference-scaled NACV, the multi-valued character of NACVs in presence of CIs makes them unsuited to be the target property of the inherently single-valued output of NNs. The dyad matrix, always being single-valued, is the property that should be learned.}

 One way in which our scheme could be improved is in replacing the rotation matrices depending on local frames by a more automatized procedure, so that any and all atoms can leave the local environment of each other.  One approach would be to redeem the idea of using a Hessian~\cite{westermayr2020combining} indirectly as an atom-wise descriptor in the dyad calculation of our approach.  This will be investigated in future work.

As it stands, the main remaining source of error in NACV prediction using the dyad method lies in a good prediction of the energy gaps used in Eq.~\eqref{eq:NACVHF}. The cusped character of adiabatic PESs around CIs makes them hard to reproduce for NNs. { Still, our dyad results show a clear improvement over CFA even with the current way of predicting energy gaps, as shown in detail in the figures in the Supplementary Material}. Recent works aiming to obtain ML-predicted eigenvalues from diagonalization of NN-learned smooth Hamiltonian representations of the system, either related to a diabatization scheme\cite{cignoni2024electronic} or effective hamiltonians\cite{westermayr2021physically}, showed very promising results. Our dyad method can be combined with any approach to obtain energies, which makes further investigations of a variety of schemes combined with DeePMD descriptors possible. {In particular, to overcome the challenge in learning the sharply varying energies near a CI, the characteristic polynomial method of Ref.~\cite{wang2023machine} to learn intersection seams is very appealing, and avoids issues arising in alternative approaches based on diabatization, such as their generally ill-defined character~\cite{mead1982conditions}, and the need for providing user-defined constraints to the NN~\cite{shu2021permutationally}. 
  While the method of Ref.~\cite{wang2023machine} only provides PESs, its use together with the dyad method would allow the two methods to complement each other.}

\section{Supplementary Material}
The Supplementary Material consists of:

 (1) a pdf document containing details of several aspects of the work, including how local frames are determined, the comparison and analysis of both the CFA and dyad methods with the 2-body embedding descriptor and our new proposed embedding descriptor, and a detailed analysis of the performance of the CFA and dyad methods near the S$_1$/S$_0$ and S$_2$/S$_1$ conical intersections, and
 
 (2) a tar file consisting of codes and data that were used in this work. 
 
\acknowledgments{
Financial support from the National Science Foundation Award CHE-2154829 (NTM),  the Computational Chemistry
Center: Chemistry in Solution and at Interfaces funded by the
U.S. Department of Energy, Office of Science Basic Energy
Sciences, under Award DE-SC0019394 as
part of the Computational Chemical Sciences Program (LD) are gratefully acknowledged.}

\bibliography{ref_na.bib}

\newpage
\clearpage



\onecolumngrid

\pagebreak
\widetext
\begin{center}
\textbf{\large Supplementary Material: Exciting DeePMD}
\end{center}
\setcounter{equation}{0}
\setcounter{figure}{0}
\setcounter{table}{0}
\setcounter{page}{1}
\setcounter{section}{0}
\makeatletter
\renewcommand{\theequation}{S\arabic{equation}}
\renewcommand{\thefigure}{S\arabic{figure}}
\renewcommand{\thetable}{S\arabic{table}}
\renewcommand\thepage{S\arabic{page}}
\renewcommand{\thesection}{S-\Roman{section}}
\renewcommand{\bibnumfmt}[1]{[S#1]}
\renewcommand{\citenumfont}[1]{S#1}


\section{Local frames}
\begin{figure}[hbtp]
\begin{overpic}[width=.5\linewidth,right]{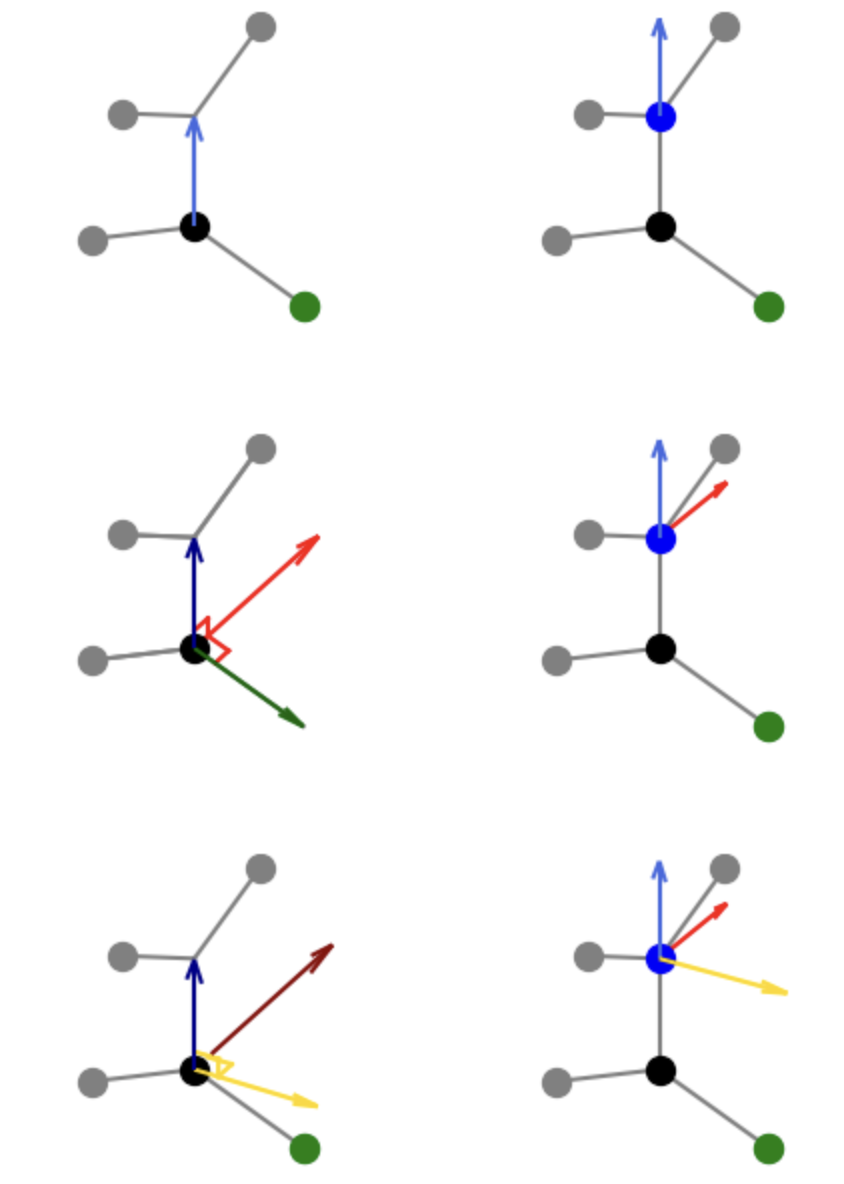}
 \put (8,60) {\Large I. \; $\mathbf{e_z} \propto \textbf{CN} $}
 \put (8,35) {\Large II. \;$\mathbf{e_y} \propto \mathbf{e_z} \times \textbf{CH}_1 $}
  \put (8,10) {\Large III. \;  $\mathbf{e_x} \propto \mathbf{e_y} \times \mathbf{e_z} $}
 \put (62,59) {\color{RoyalBlue}\Large \textbf{CN}}
 \put (58,36) {\color{Blue}\Large $\mathbf{e_z}$}
 \put (62,24) {\color{OliveGreen}\Large \textbf{CH}$_1$}
 \put (58,10) {\color{Blue}\Large $\mathbf{e_z}$}
 \put (66,15) {\color{BrickRed}\Large $\mathbf{e_y}$} 
 \put (87,53) {\large C}
 \put (90,60) {\color{blue} \large N}
 \put (96,49) {{\color{ForestGreen} \large  H$_1$}}
 \put (81,65) {\large H$_2$}
 \put (94,70) {\large H$_3$}
 \put (79,55) {\large H$_4$}
 \put (86,69) {{\color{RoyalBlue}\Large $\mathbf{e_z}$}}
  \put (86,45) {{\color{RoyalBlue}\Large $\mathbf{e_z}$}}
 \put (86,19) {{\color{RoyalBlue}\Large $\mathbf{e_z}$}}
  \put (94,40) {{\color{red}\Large $\mathbf{e_y}$}}
  \put (94,16) {{\color{red}\Large $\mathbf{e_y}$}}  
  \put (97,11) {{\color{Goldenrod}\Large $\mathbf{e_x}$}}    
\end{overpic}     
    \caption{Construction steps for the local frame shared by atoms C, N and H$_1$. Each row shows the construction of one of the basis vectors.}
    \label{fig:frame}
\end{figure}

Here in Figure \ref{fig:frame} we show how the local frame used by atom C, N and H$_1$ atom-wise descriptors is built. We need to provide 3 orthogonal basis vectors to define it: First, $\mathbf{e_z}$ is taken collinear to the vector \textbf{CN}. Second, $\mathbf{e_y}$ has its direction along the vector product of \textbf{CN} and \textbf{CH}$_1$. Finally, $\mathbf{e_x}$ is obtained as the vector product of $\mathbf{e_y}$ with $\mathbf{e_z}$. For atoms H$2$ to H$4$ the procedure is exactly the same, but instead of using H$_1$ the H atom considered is used. Thus, while C, N, and H1 share the same local frame, the other H atoms do not.

\newpage
\section{Analysis of shortcomings of CFA and embedding descriptors}

To complete the comparison of CFA and dyad method done in the main text, we show here the performance of alternate ML schemes combining the 2-body embedding descriptor with the dyad method and our new embedding descriptor with the CFA method. These results demonstrate the overall superiority of our dyad approach, and also justify our choice to focus the main discussion on the combination of CFA with usual DeePMD embedding and the combination of our dyad predictor with the new descriptor. We then discuss how these results could be anticipated from the structure of CFA and dyad predictors.

\begin{figure}[hbtp]
    \centering
    \includegraphics[width=.75\linewidth]{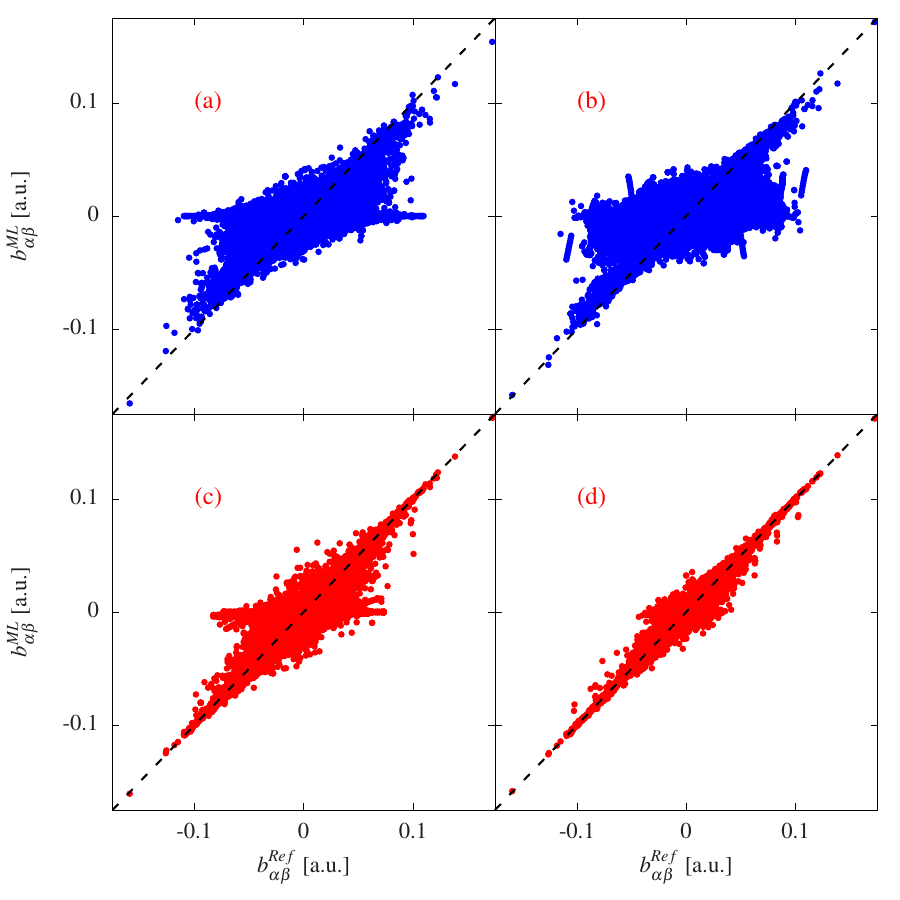}
    \caption{Comparison of prediction accuracy for different descriptor and predictor combinations: 
    a) CFA predictor with 2-body embedding. 
    b) CFA predictor with our new embedding.
    c) Dyad predictor with 2-body embedding. 
    d) Dyad predictor with our new embedding. Only the $y$ components of vectors are shown.}
    \label{fig:CFvsDyad}
\end{figure}

First, in Figure \ref{fig:CFvsDyad} we compare the prediction capabilities of different combinations of descriptors and predictors for  the $y$ components of $\mathbf{b}_{\alpha\beta}$ over the training set. All hyperparameters have identical values to what is reported in the main manuscript.  The top panels in the figure correspond to the CFA predictor, while the bottom panels correspond to the dyad. These are used in conjunction with the DeePMD 2-body embedding in the left panels and the new embedding descriptor in the right panels. We also report MAE and RMSE of each method for each cartesian component in Table \ref{tab:ML_vs_QC}. 

\begin{table}[!htbp]
\centering
\caption{Mean absolute error (MAE) and root mean square error (RMSE) on $b_{\alpha\beta}$ over the training set using each ML method shown on Fig. 2 for each cartesian component. Methods are defined by the combination of a predictor component (either CFA of Dyad) and a descriptor, with DeePMD 2-body embedding (2be) or our new embedding (de).}
\begin{tabular}{p{2.5cm} | p{1.5cm} p{1.5cm} p{1.5cm}}
\hline
\; Method &  \multicolumn{3}{c}{MAE (RMSE) [ $10^{-3}$ a.u.]} \\
\hline
{}   & \quad \quad $x$   & \quad \quad $y$    & \quad \quad $z$   \\
\quad (a) CFA-2be & \quad 5.4 (8.2) & \quad 16 (24) & \quad 8.0 (14)  \\
\quad (b) CFA-de & \quad 3.8 (5.6) & \quad 12 (20) & \quad 5.8 (11)  \\
\quad (c) Dyad-2be & \quad 3.6 (6.1) & \quad 4.9 (9.0) & \quad 4.0 (8.1)  \\
\quad (d) Dyad-de & \quad 1.9 (3.2) & \quad 2.1 (3.8) & \quad 1.9 (3.1)
       \\ \hline
\end{tabular}    \label{tab:ML_vs_QC}
\end{table}

While Figure 3 in main text matches with panels (a) and (d), respectively CFA with 2-body embedding and dyad with its corresponding embedding, here we verify that the superior performance of the dyad approach  is largely due to the choice of dyad as predictor  the CFA predictor, and that, at the same time, using the new embedding descriptor with the dyad predictor greatly improves its accuracy compared with using the 2-body embedding. 
 Comparing panels (a) and (c) shows that the dyad predictor achieves an improvement over the CFA predictor even when both are using 2-body embedding descriptor. Interestingly, panel (c) shows a pattern in appearance similar to (a), with a subset of components being predicted as essentially $0$ even if the reference values are distributed over the range $[-0.1;0.1]$ in atomic units. Their origin is nevertheless completely different: While the null values in panel (a) are due to the theoretical inability of the CFA to predict non-zero out-of-plane components for a planar molecular geometry, as discussed in the text,
the null values in panel (c) are caused by the inadequacy of the atom-wise 2-body embedding when used to predict blocks of $\mathbf{\Gamma_{\alpha\beta}}$, as we show below.

\begin{figure}[hbtp]
    \centering
\begin{overpic}[width=0.5\linewidth]{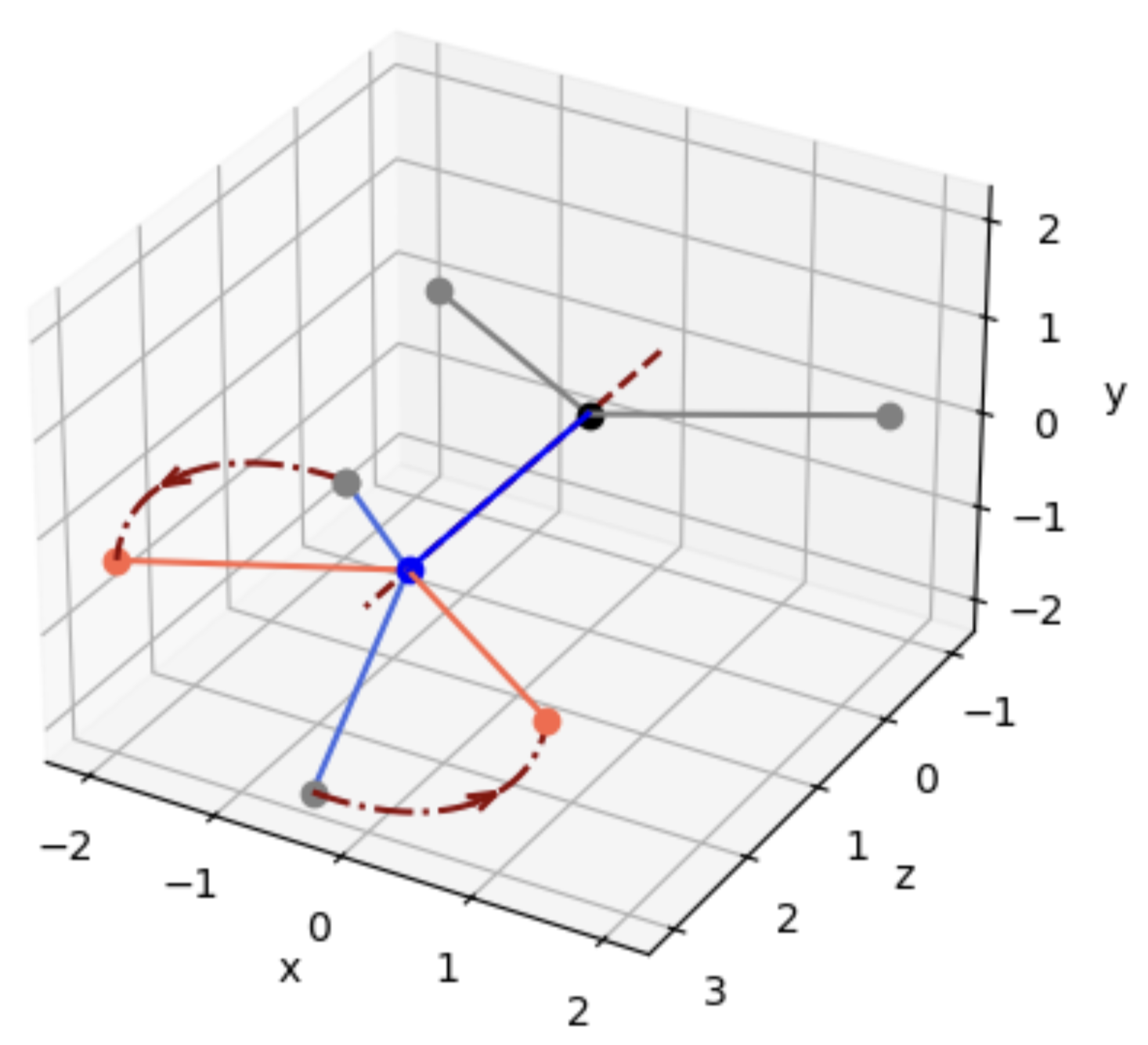}
 \put (50,50) {C}
 \put (40,40) {\color{blue}N}
 \put (70,60) {H$_1$}
 \put (20,25) {H$_2$}
 \put (25,55) {H$_3$}
 \put (40,70) {H$_4$}
\end{overpic}     
    \caption{Plot of the two geometries compared and the rotation that transforms one into the other.}
    \label{fig:Geom}
\end{figure}

Consider the two particularly geometries depicted on Figure \ref{fig:Geom}. They differ by the position of atoms H2 and H3, showed in grey with blue bonds for one geometry $\mathbf{R}$ and in red with red bonds for the other $\tilde{\mathbf{R}}$. 
 In each geometry, the C, N, H2, and H3 atoms all lie in the same plane, with the C-N axis bisecting the line between H2 and H3. 
The second geometry is obtained by a rotation of these two hydrogens around the C-N axis  by 90 degrees as depicted by the dark-red arrows  but our argument below holds for rotation of any angle about this axis. 

\begin{figure}[hbtp!]
    \centering
     \begin{subfigure}[t]{0.3\textwidth}
         \centering
\begin{overpic}[width=\linewidth]{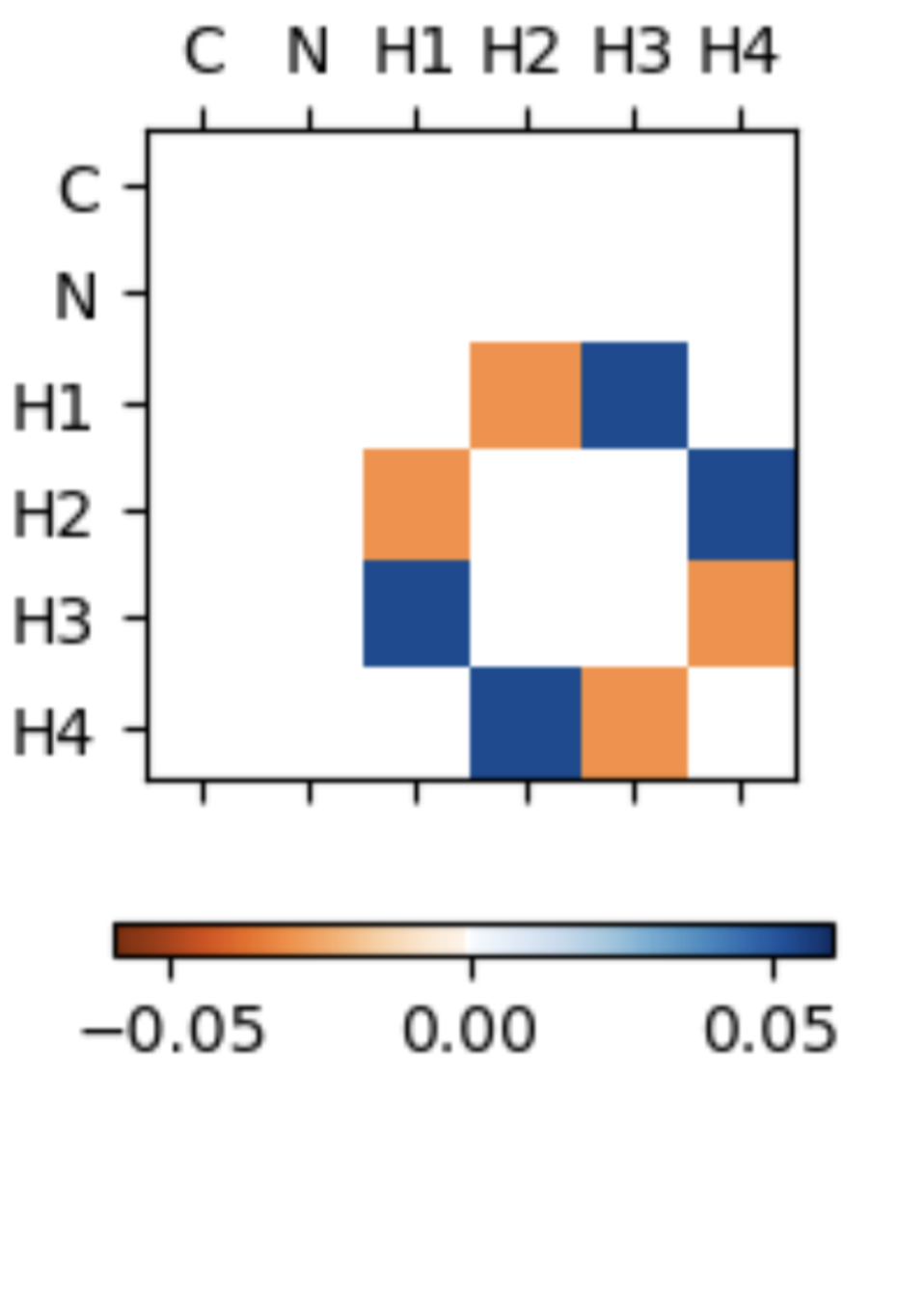}
 \put (10,5) {\large $\displaystyle (\Delta S)_{ij} = s(r_{ij})-s(\tilde{r}_{ij})$}
\end{overpic}         
    \caption{Plot of the difference of generalized distance matrices for the initial and rotated geometries.}
    \label{fig:DeltaS}
     \end{subfigure}
     \hfill
     \begin{subfigure}[t]{0.65\textwidth}
         \centering
\begin{overpic}[width=\linewidth]{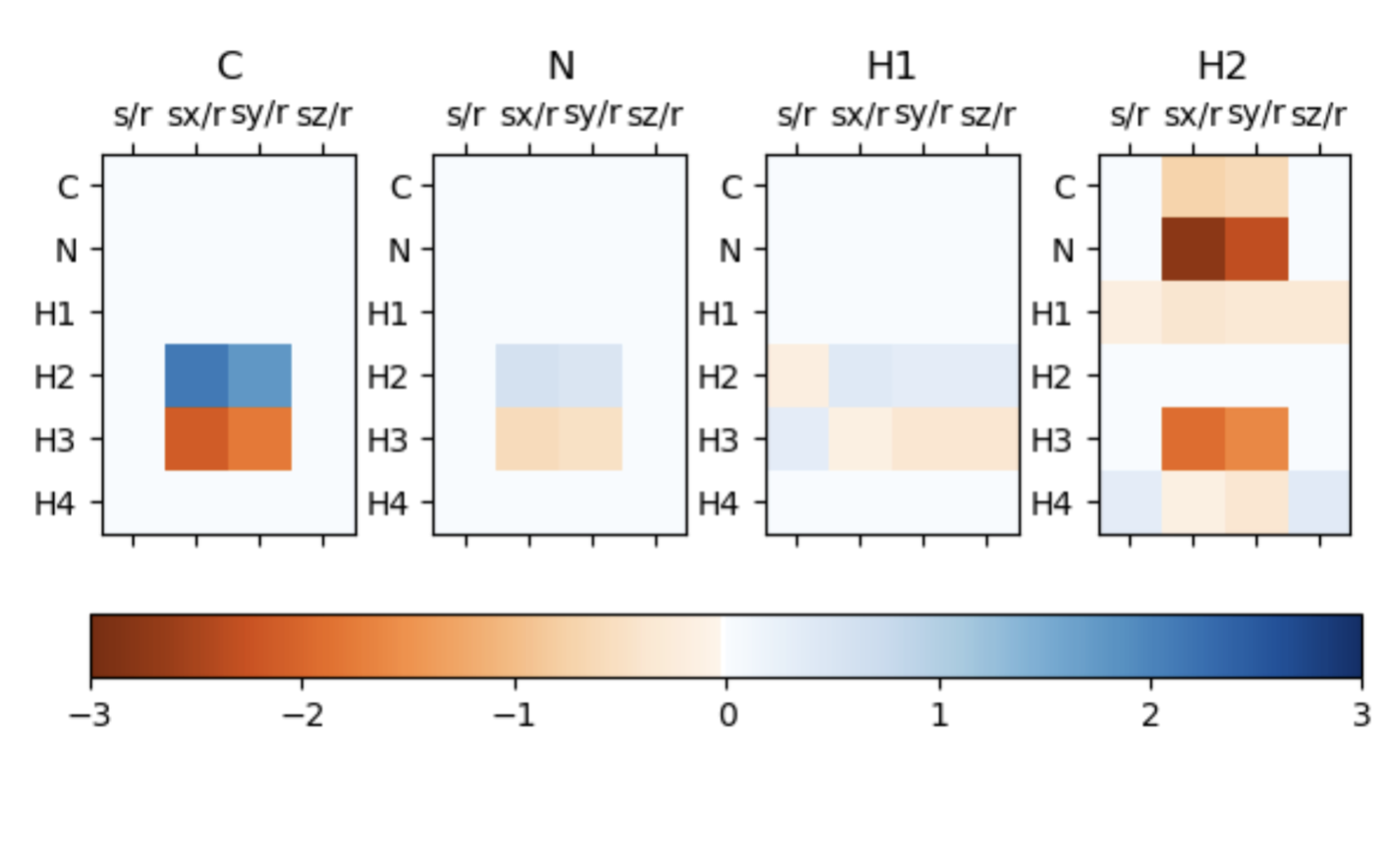}
 \put (35,3) {\large $\displaystyle (\Delta \mathcal{R}^i)_{j} = (\mathcal{R}^i)_{j} - (\tilde{\mathcal{R}}^i)_{j} $}
\end{overpic}          
    \caption{Plot of the difference of generalized coordinate matrices for the initial and rotated geometries.}
    \label{fig:DeltaR}
     \end{subfigure}
    \caption{Effect of rotation on generalized coordinate and distance matrices.}
    \label{fig:DeltaSR}
\end{figure}

Our goal is to show that the atom-wise embedding of C and N atoms cannot distinguish these two molecular geometries. We recall the definition of 2-body embedding:
\begin{equation}
\label{eq:2bD}
    \mathcal{D}^i = \frac{1}{N_c^2} \left( \mathcal{G}^i \right)^T  \mathcal{R}^i  \left( \mathcal{R}^i \right)^T \; \mathcal{G}^i_<
\end{equation}
where the embedding NN, denoted $\mathcal{N}_{e,2}$ is fed with radial information only:
\begin{equation} 
\label{eq:2bG}
    \left( \mathcal{G}^i \right)_j = \mathcal{N}^{p_{ij}}_{e,2}\big(s(r_{ij})\big)
\end{equation}
Clearly, the rotation described above does not change the distance of any atom to N nor to C, as illustrated in Fig.~\ref{fig:DeltaS},  which plots the difference between $s(r_{ij})$ and $s(\tilde{r}_{ij})$, the tilde denoting values obtained for the rotated geometry. As seen from Eqs \eqref{eq:2bD} and \eqref{eq:2bG}, in each atom-wise descriptor $\mathcal{D}^i$ the NN $\mathcal{N}^{p_{ij}}_{e,2}$ is only fed with one row of $(S)_{ij}=s(r_{ij})$. Hence, we deduce the output of embedding NN $\mathcal{G}^i$ for C ($i=1$) and N ($i=2$) does not distinguish the two geometries while for all H atoms $\mathcal{G}^i$ will be affected by the rotation, as seen in Fig.~\ref{fig:DeltaS}.

  But $\mathcal{D}^i$ is also affected by the change of $\mathcal{R}^i\big(\mathcal{R}^i\big)^T$ induced by the rotation (Eq.~(\ref{eq:2bD})), even if the embedding NN for C and N is oblivious to the rotation. For all atoms, $\tilde{\mathcal{R}}^i \neq \mathcal{R}^i$ as can be seen from Figure \ref{fig:DeltaR} where we plot their difference for $i=1...4$. However, this distinguishability is lost when the angular information is contracted by the matrix-vector products of $(\mathcal{G}^i)^T$ and $\mathcal{R}^i$ for $i=1$ and 2, as we will next show.

\begin{figure}[hbtp!]
    \centering
     \begin{subfigure}[b]{0.6\textwidth}
         \centering
    \begin{overpic}[width=\linewidth]{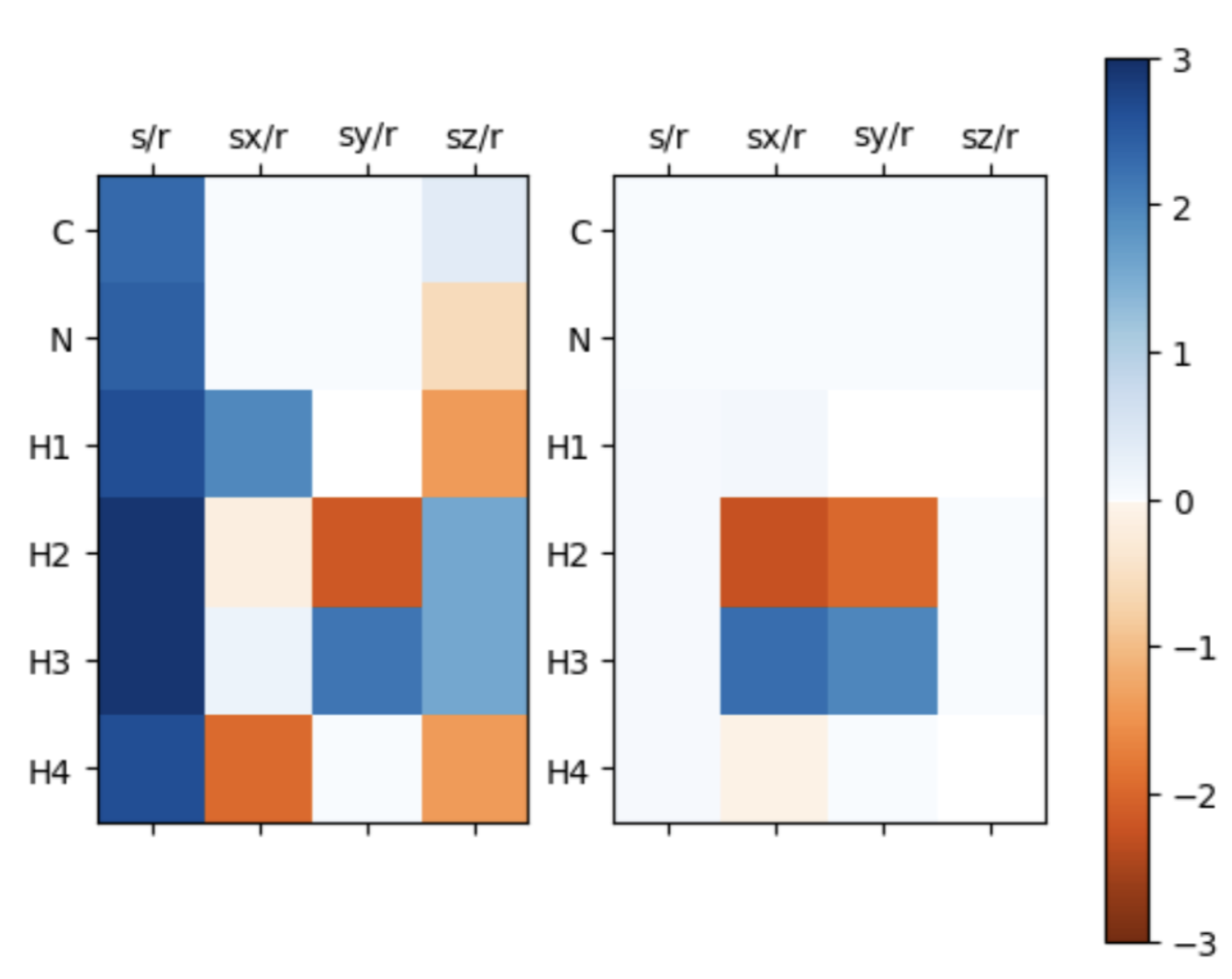}
 \put (15,75) {\large $\displaystyle s(\mathbf{r}_{i})^T . \,\mathcal{R}^i$}
 \put (48,75) {\large $\displaystyle s(\mathbf{r}_{i})^T . \,\mathcal{R}^i - s(\tilde{\mathbf{r}}_{i})^T . \,\tilde{\mathcal{R}}^i$}
\end{overpic}          
         \caption{Left panel: Plot of the contraction of generalized distance and coordinate matrices for the initial geometry.
    Right panel: Difference between initial and rotated geometries.}
         \label{fig:HalfD}
     \end{subfigure}
     \hfill
     \begin{subfigure}[b]{0.35\textwidth}
         \centering
    \begin{overpic}[width=.8\linewidth]{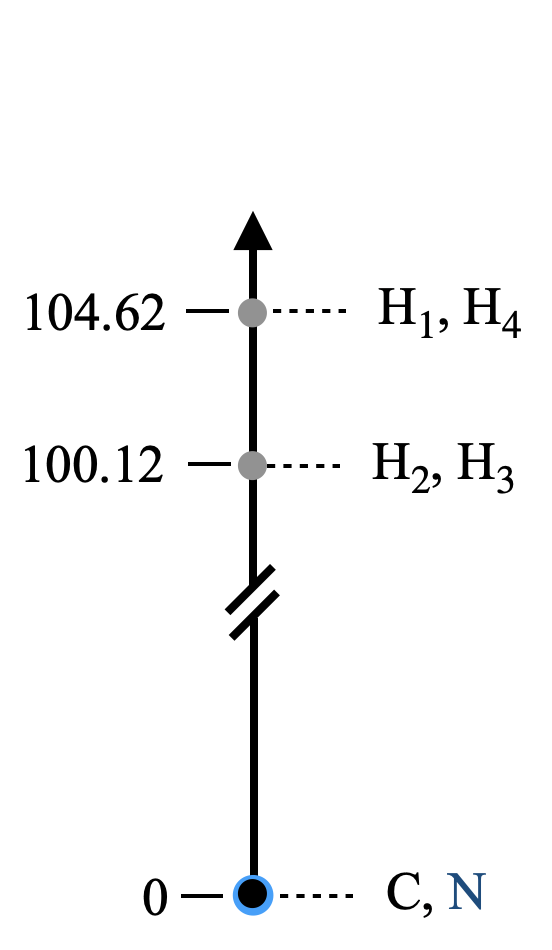}
 \put (7,90) {\large $\displaystyle \Delta \mathcal{D}^i = \big|\big| \mathcal{D}^i - \tilde{\mathcal{D}}^i \big|\big|$}
\end{overpic}          
         \caption{Measure of difference on atom-wise 2-body embedding descriptors for the initial and rotated geometries.}
         \label{fig:DeltaD}
     \end{subfigure}
    \caption{Effect of rotation on 2-body embedding descriptors.}
    \label{fig:D}
\end{figure}

As before, without loss of generality we replace $(\mathcal{G}^i)_j$ with $s(r_{ij})$. The matrix-vector product yields a 4-component vector for each atom-wise descriptor, and those are given for the initial geometry on left panel of Figure \ref{fig:HalfD}. We see that for C and N atoms, the $x$ and $y$ components are strictly 0 as the contributions of H2 exactly compensates that of H3 because of symmetry (for this example, the same is true of H1 and H4). This fact is not affected by any rotation of the H pair around the C-N bond. Thus, $(\mathcal{G}^i)^T.\mathcal{R}^i$ does not detect the rotation for C and N. This is shown on the right panel, plotting the difference between those partial descriptors for the initial and rotated geometries.

As the full 2-body embedding descriptor is obtained by taking the matrix-product of $(\mathcal{G}^i)^T \mathcal{R}^i$ with its transpose, it will inherit its invariances. This is illustrated on Figure \ref{fig:DeltaD} which shows the distance between descriptors of initial and rotated geometries for all atoms. The distance is computed as:

\begin{equation}
    \big|\big| \mathcal{D}^i - \tilde{\mathcal{D}}^i \big|\big| = \sum_{k}^M \sum_l^{M_<} \big|(\mathcal{D}^i)_{kl} - (\tilde{\mathcal{D}}^i)_{kl} \big|
\end{equation}

By comparing Figures 1 and 2 in the main text, one can see that in CFA all atom-wise descriptors are combined to compute the fictitious anti-derivative of $\mathbf{b}_{\alpha\beta}$. Thus, it is sufficient that the H atoms spot the rotation. The situation is different in the dyad prediction method as the C and N embedding descriptors are used exclusively to predict diagonal blocks of $\Gamma_{\alpha\beta}$ and thus need to hold more information.

 Returning to Fig.~\ref{fig:CFvsDyad}, comparison of panels (a) and (b) show that using the more complex dyad-adapted atom-wise embedding with the CFA predictor does not lead to an improvement comparable to that of the dyad predictor on panel (d). This could be anticipated since the way all embedding features are summed over in the CFA approach (see Figure 1 of main text) makes the wealth of information of our new descriptor quite redundant. Our efforts to not contract angular information and enrich each atom-wise descriptor with more information offer no advantage when fed to the CFA predictor.

\newpage
\section{Performance in the vicinity of the conical intersections}

While the previous section analyzed the quality of ML predictions over the whole dataset, we expect the vicinity of conical intersections (CIs) is the most crucial for reproducing the correct dynamics. In this section, we focus on how well the dyad and CFA methods can match ab initio data around the S1/S0 and S2/S1 CIs.  To do so, we use two subparts of the training dataset as scans of CI neighborhoods along effective one-dimensional paths. We will also discuss the effect of using the L2 loss or the phaseless loss during training for the CFA method.

\subsection{S2/S1 CI}

\begin{figure}[hbtp]
    \centering
    \includegraphics[width=0.6\linewidth]{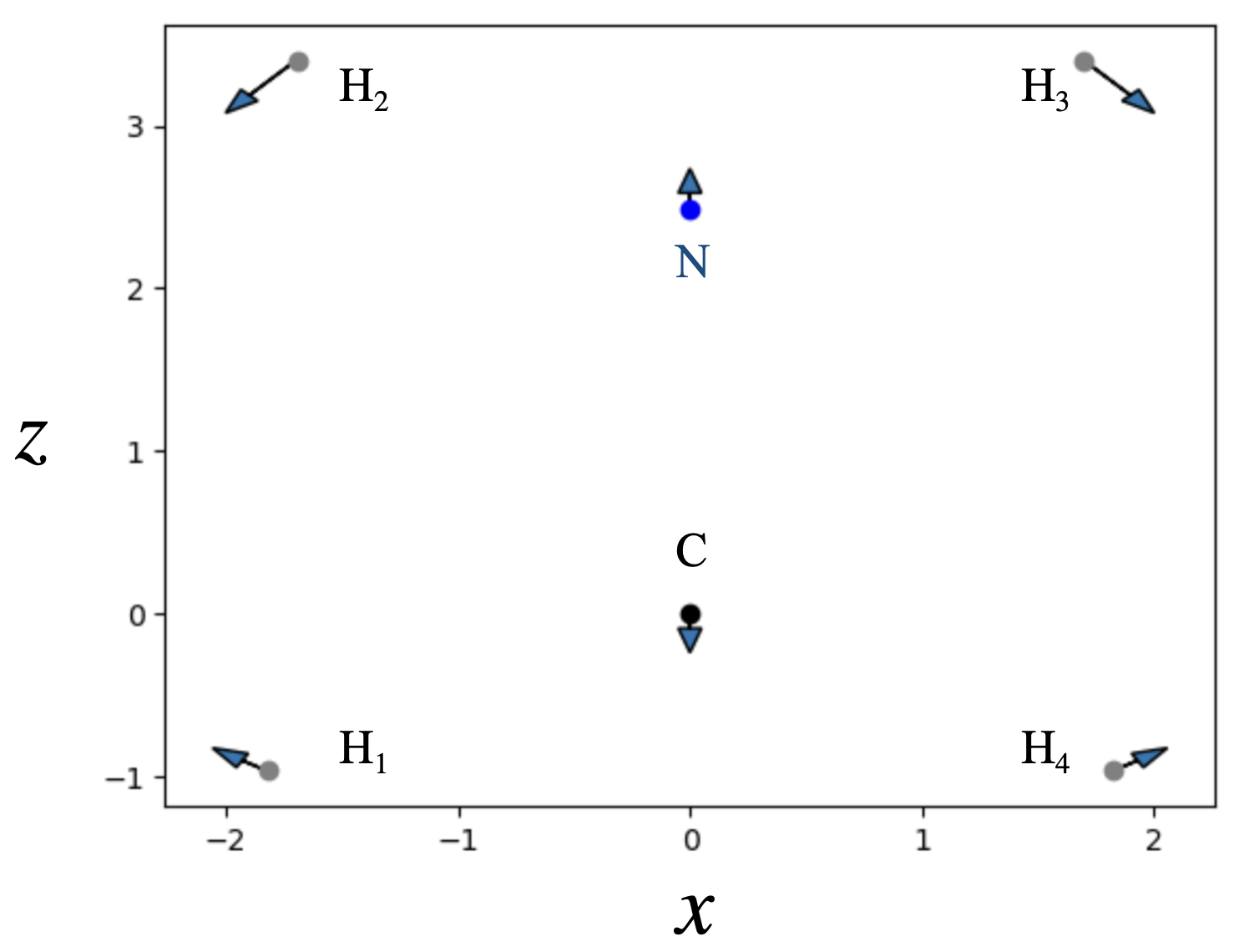}
    \caption{Geometries in subpart of dataset used to study the vicinity of S$_1$/S$_2$ CI. All geometries lie in the $y=0$ plane, thus only the $x$ and $z$ coordinates are shown. The dots show the geometry 3840, the arrows show the deformation of the molecule induced by going through geometries 3840 to 3875.}
    \label{fig:S2S1mov}
\end{figure}

\begin{figure}[hbtp!]
    \centering
    \includegraphics[width=0.6\linewidth]{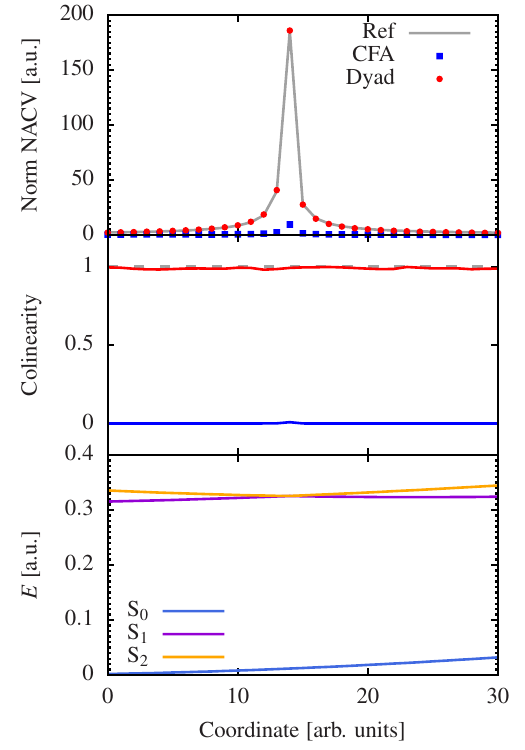}
    \caption{Performance of ML methods for NACV prediction in the vicinity of the S$_1$/S$_2$ CI. Top panel: Norm of NACV between states 1 and 2 compared to reference values (exact energy gaps are used to compute $\mathbf{d}_{\alpha\beta}$ from $\mathbf{b}_{\alpha\beta}$). Middle Panel: Colinearity of predicted and reference NACVs. Bottom panel: Exact energies for the 3 states.}
    \label{fig:S2S1NAC}
\end{figure}

The S2/S1 CI is scrutinized by considering geometries labelled 3840 to 3875 in the dataset (provided in the supplementary material of Ref. 16 of the main text). Movement along the effective coordinate corresponds to an elongation of the CN bond together with an opening of the angle of the H atoms with the CN axis. This is illustrated in Figure \ref{fig:S2S1mov}. The potential energies along the coordinate are plotted in last panel of Figure \ref{fig:S2S1NAC}.

To ease the analysis, we will use two measures of the quality of ML prediction: the magnitude of true and learned NACVs, and  the vector collinearity of the prediction with the reference values, i.e. 
$ \displaystyle \frac{\mathbf{d}_{\alpha\beta}^{\text{Ref}}.\mathbf{d}_{\alpha\beta}^{\text{ML}}}{|\mathbf{d}_{\alpha\beta}^{\text{Ref}}| \; |\mathbf{d}_{\alpha\beta}^{\text{ML}}|}  $.

In an effort to isolate the error of the predicted $\mathbf{b}_{\alpha\beta}$ from that of the predicted energy gaps appearing in the denominator of the NACV, we instead divide the predicted numerator ${\bf b}_{\alpha\beta}$ by the exact energy gaps.
The norm of NACVs predicted by the dyad method (red dots) and by CFA (blue squares) are plotted together with the reference values (grey line) on top panel of Figure \ref{fig:S2S1NAC}. The CFA prediction strikingly fail to reproduce the magnitude of the reference values while the dyad prediction is quantitative. The collinearity of the ML predictions with the reference values is plotted in the middle panel of Figure \ref{fig:S2S1NAC} with same colors as top panel.

Again, the dyad method is seen to reproduce the direction of the true NACVs with quantitative accuracy, the collinearity being always above 0.975, while the CFA prediction is almost orthogonal to the reference. There is no noticeable difference between results obtained using L2 loss or phaseless loss for training of CFA ML. The NACV values are essentially perpendicular to the $xz$ plane for these geometries, so, in light of the CFA being unable to predict non-zero vectors perpendicular to the molecular plane, this failure is expected. 
On the other hand, the quantitative collinearity of the dyad prediction means it not only correctly predicted the large $y$ components, but also the almost 0 components along $x$ and $z$. This shows the dyad method performs well even when a combination of small and large components is involved. 

\begin{figure}[hbtp!]
    \centering
    \includegraphics[width=0.6\linewidth]{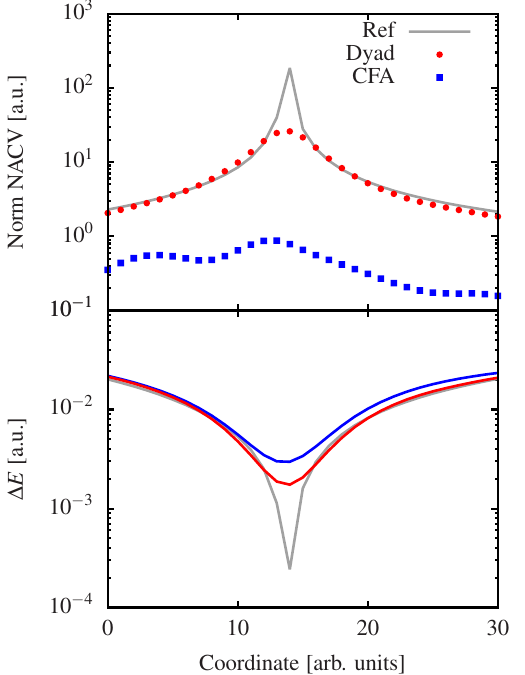}
    \caption{Performance of ML methods for NACV prediction in the vicinity of the S$_1$/S$_2$ CI. Top panel: Norm of NACV between states 1 and 2 compared to reference values (NN predictions of energy gaps are used to compute $\mathbf{d}_{\alpha\beta}$ from $\mathbf{b}_{\alpha\beta}$). Bottom panel: Reference and predicted energy gaps between states 2 and 1.}
    \label{fig:S2S1tNAC}
\end{figure}

To see the effect of energy predictions, we compare the ML energy gaps $E_{S_2} -E_{S_1}$ to the reference values (grey line) on the bottom panel of Figure \ref{fig:S2S1tNAC}. The energy learning is not affected by the NACV prediction method (dyad of CFA) and thus the difference between the two predictions is due to the stochasticity of the training. We see that the two predictions of the energy gaps agree well with the reference overall, except when the energy gap becomes small enough that it approaches the overall error in the ML energy prediction.
 The energy gaps around a CI can become arbitrarily small, thus lower than any reasonable accuracy threshold.  In the top panel of Figure \ref{fig:S2S1tNAC} we report the norm of the predicted NACVs when $\mathbf{b}_{\alpha\beta}$ is divided by the ML-predicted energy gaps. Note the collinearity is unaffected by the energy gaps and thus is not shown again. Comparison with top panel of Figure~\ref{fig:S2S1NAC} shows the learned energy gaps to be the main remaining source of error for the dyad method, while it does not prevent it to yield predictions of much better quality than the CFA approach.

\clearpage
\subsection{S1/S0 CI}

\begin{figure}[hbtp!]
    \centering
    \includegraphics[width=0.6\linewidth]{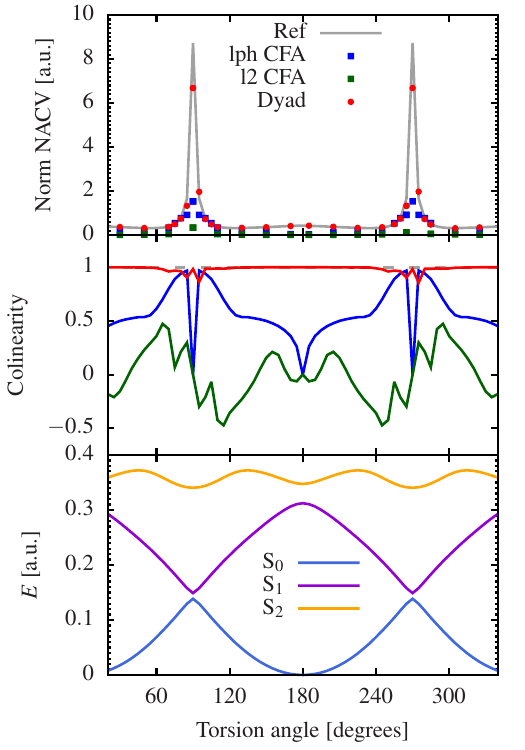}
    \caption{Performance of ML methods for NACV prediction in the vicinity of the S$_0$/S$_1$ CI. CFA results were obtained using the L2 loss for NACV during training (l2 CFA, green squares) and using the phase-less loss (lph CFA, blue square). Top panel: Norm of NACV between states 0 and 1 compared to reference values (exact energy gaps are used to compute $\mathbf{d}_{\alpha\beta}$ from $\mathbf{b}_{\alpha\beta}$). Middle Panel: Colinearity of predicted and reference NACVs. Bottom panel: Exact energies for the 3 states.}
    \label{fig:S1S0NAC}
\end{figure}

The S1/S0 CI is scrutinized by considering geometries 3010 to 3043 and 3078 to 3044 (notice the order is reversed). With this reordering, these points sample the same motion as that depicted on Figure \ref{fig:Geom}: C, N, H$_3$ and H$_4$ stay fixed in the $xz$ plane while H$_2$ and H$_3$ rotate along the CN axis. The bottom panel of Figure \ref{fig:S1S0NAC} shows how the potential energies change with this motion.

We compare the norm of dyad and CFA predictions to reference NACV values in the top panel of Figure \ref{fig:S1S0NAC}. Again, exact energy gaps are used to scale $\mathbf{b}_{\alpha\beta}$ to isolate the error of the approach used to predict $\mathbf{b}_{\alpha\beta}$ from errors energy prediction. The colinearity of the ML results with the true vector (middle panel) shows the  quality of CFA predictions depends on the loss function used to train the NACVs, even if the dataset is already phase-corrected. This shows that the multi-valuedness of the NACV vector field, a property that the phase-correction cannot remove, is impeding the capability of the CFA method to learn NACVs. The phaseless loss gives better results as it is able to flip the sign of NACVs in the [90:270] torsion angle range. While it makes the ML prediction better at those geometries, it is clear the phaseless loss does not enable us to learn the true NACV field well. Moreover, the CFA method cannot predict the NACVs faithfully away from the CI, neither in magnitude nor in orientation, no matter what loss function is used for NACVs. As expected, as the molecule comes close to a planar geometry (torsion angle of 180 degrees), limitations of the CFA approach prevent it from predicting the out-of-plane component and thus both the norm and collinearity collapse to 0. These geometries however are far away from the CI which lies at a torsion angle of 90 degrees.  Using phase-less loss, CFA does better as we get close to that angle, but then suffers a dramatic decrease in accuracy as the vicinity of the CI is reached. Further analysis showed that this is due to the $x$ components of NACVs being wrongly predicted to be 0 by the CFA when the 90 degree angle is reached. The molecule acquires a C$_{2\text{v}}$ symmetry with both a $y=0$ symmetry plane and a $x=0$ symmetry plane. The unphysical constraint of being a symmetry-invariant conservative field that is enforced by the CFA is therefore seen to have a deleterious effect even in the non-planar molecule case. On the other hand, predictions from the dyad method are faithful to the reference NACV for all torsion angles, both in magnitude and orientation.

\begin{figure}[hbtp!]
    \centering
    \includegraphics[width=0.6\linewidth]{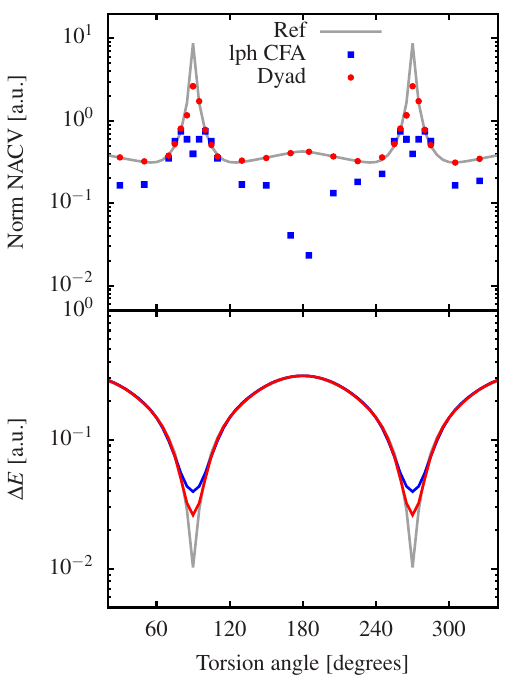}
    \caption{Performance of ML methods for NACV prediction in the vicinity of the S$_0$/S$_1$ CI. Top panel: Norm of NACV between states 0 and 1 compared to reference values (NN predictions of energy gaps are used to compute $\mathbf{d}_{\alpha\beta}$ from $\mathbf{b}_{\alpha\beta}$). Bottom panel: Reference and predicted energy gaps between states 0 and 1.}
    \label{fig:S1S0tNAC}
\end{figure}

In Figure \ref{fig:S1S0tNAC}, ML-predicted energy gaps $E_{S_1} - E_{S_0}$ are compared to reference values on bottom panel while their impact on ML prediction for NACVs can be assessed from top panel, plotting the norm of couplings. Again, it is seen that the norm of NACVs predicted with the dyad method are the closest match to the reference values, and the strong improvement it offers compared to CFA approach is not made useless by errors in the predicted energy gaps.

\end{document}